\documentclass[aps,prl,twocolumn,superscriptaddress,showpacs,showkeys]{revtex4-1}

\usepackage{graphicx}
\usepackage{amsmath, amsxtra, amssymb, mathtools, isomath, nccmath, xspace, siunitx}
\usepackage{color}
\usepackage[utf8]{inputenc}
\usepackage[british]{babel}
\usepackage{hyperref}
\usepackage{sidecap}
\usepackage{bm}
\usepackage{multirow}

\newcommand{\ket}[1]{\ensuremath{\lvert #1 \rangle}\xspace}%
\newcommand{\parallelsum}{\mathbin{\!/\mkern-5mu/\!}}

\newcommand{\NaK}{$^{23}$Na$^{40}$K }

\sidecaptionvpos{figure}{h}

\begin{document}
\title{\bf{Extending rotational coherence of interacting polar molecules\\ in a spin-decoupled magic trap}} 

\author{Frauke Seeßelberg}
\affiliation{Max-Planck-Institut f\"{u}r Quantenoptik, 85748 Garching, Germany}

\author{Xin-Yu Luo}
\email[]{xinyu.luo@mpq.mpg.de}
\affiliation{Max-Planck-Institut f\"{u}r Quantenoptik, 85748 Garching, Germany}

\author{Ming Li}
\affiliation{Department of Physics, Temple University, Philadelphia, Pennsylvania 19122, USA}

\author{Roman Bause}
\affiliation{Max-Planck-Institut f\"{u}r Quantenoptik, 85748 Garching, Germany}

\author{Svetlana Kotochigova}
\affiliation{Department of Physics, Temple University, Philadelphia, USA}

\author{Immanuel Bloch}%
\affiliation{Max-Planck-Institut f\"{u}r Quantenoptik, 85748 Garching, Germany}
\affiliation{Fakult\"{a}t f\"{u}r Physik, Ludwig-Maximilians-Universit\"{a}t M\"{u}nchen, 80799 M\"{u}nchen, Germany}%

\author{Christoph Gohle}
\affiliation{Max-Planck-Institut f\"{u}r Quantenoptik, 85748 Garching, Germany}

\date{\today}


\begin{abstract}
Superpositions of rotational states in polar molecules induce strong, long-range dipolar interactions.
Here we extend the rotational coherence by nearly one order of magnitude to 8.7(6)~ms in a dilute gas of polar \NaK molecules in an optical trap.
We demonstrate spin-decoupled magic trapping, which cancels first-order and reduces second-order differential light shifts.
The latter is achieved with a dc electric field that decouples nuclear spin, rotation, and trapping light field.
We observe density-dependent coherence times, which can be explained by dipolar interactions in the bulk gas.
\end{abstract}

\maketitle

Interacting particles with long coherence times are a key ingredient for entanglement generation and quantum engineering. 
Cold and ultracold polar molecules \cite{Ni2008, Shuman2010, Takekoshi2014, Molony2014, Park2015,Guo2016,Prehn2016,Rvachov2017,Seesselberg2018,Anderegg2018,Collopy2018} are promising systems for exploring such quantum many-body physics with long-range interactions \cite{Moses2016a, DeMarco2018} due to their strong and tunable electric dipole moment and long single-particle lifetime \cite{Chotia2012,Yan2013a}. 
The manipulation of their rich internal degrees of freedom has been studied for different molecular species \cite{Ospelkaus2010, Will2016, Guo2018, Blackmore2018}.
First observations include ultracold chemistry and collisions \cite{Ni2010, Ye2018}.  
Nuclear spin states in the rovibronic ground state further promise exciting prospects for quantum computation due to their extremely long coherence times \cite{Park2017}.

Rotation is a particularly appealing degree of freedom for molecules because it is directly linked to their dipolar interactions.
It can be manipulated by microwave (MW) fields and superpositions of rotational states with opposite parity exhibit an oscillating dipole moment with a magnitude close to the permanent electric dipole moment $d_0$. 
Consequently, using rotating polar molecules has been proposed for quantum computation \cite{DeMille2002}, to emulate exotic spin models \cite{Peter2012} or to create topological superfluids \cite{Yao2013}.

In order to make use of the rotational transition dipole in a spatially inhomogeneous optical trap, the coupling of the rotation to the trap field needs to be canceled. 
To first order this may be achieved by choosing an appropriate angle between the angular momentum of the molecule and the trapping field polarization $\bm{\varepsilon}$ \cite{Kotochigova2010} or a special trap light intensity \cite{Blackmore2018} such that the differential polarizability between rotational ground and excited states is canceled.
The trap is then referred to as ``magic".
Coherence times of about 1~ms have been achieved in bulk gases of polar molecules using these techniques \cite{Neyenhuis2012,Blackmore2018}.
However, this is much shorter than the dipolar interaction time, preventing observation of many-body spin dynamics.

The coherence time in such a magic trap is limited by the intensity dependence of the molecular polarizability, which originates from the coupling between rotation, nuclear spins, and the trapping light field. 
It has been suggested to apply large magnetic \cite{Deng2015} or electric fields \cite{Li2017} to reduce these couplings and thus simplify the polarizabilities of the involved states. 

In this work, we realize a spin-decoupled magic trap, i.e. a magic polarization angle trap with moderate dc electric fields, which simplify the hyperfine structure of the rotational transition manifold $\ket{J=0,m_J=0}\rightarrow\ket{1,0}$. 
Here, $J$ denotes the rotational quantum number and $m_J$ its projection onto the electric field axis. 
We characterize the magic trapping condition and demonstrate how the second-order light shift is related to the electric field strength. 
Using Ramsey- and spin-echo interferometry, we further study the rotational coherence time of polar molecules in a spin-decoupled magic one-dimensional (1D) lattice.
A coherence time of almost 10~ms is achieved for a dilute gas of ultracold \NaK molecules; however, we find that the coherence time decreases with increasing molecular density. 
Using a simple numerical model \cite{Hazzard2014a}, we conclude that the dipolar interaction between molecules plays a dominant role in the density-dependent decoherence.
This interaction can become as large as $h\times 50$~Hz, due to the large permanent dipole moment $d_0 = 2.72$~D of \NaK \cite{Gerdes2011}, at the highest accessible density of $6.8\times 10^{10}$/cm$^3$, comparable to the single particle dephasing.

\begin{figure}
\includegraphics{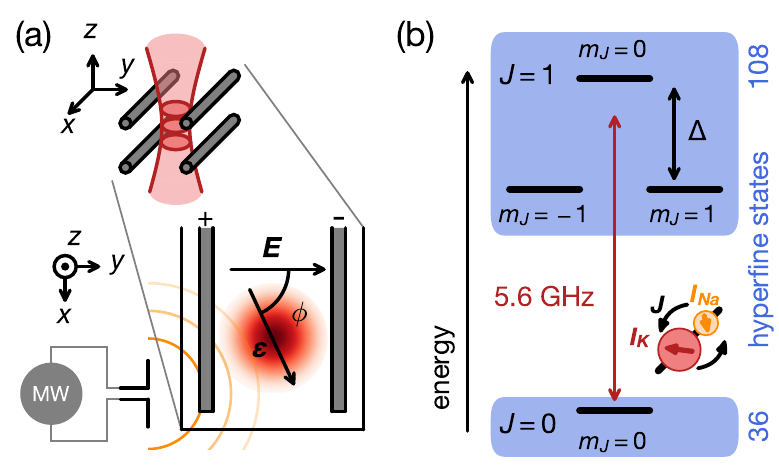}
\caption{\label{fig:schematic} 
(a) 
Schematic of the experimental setup. 
Molecules are confined to several pancake shaped optical traps (red) in the $x-y$ plane, formed by a 1D optical lattice along the $z$ axis with polarization vector $\bm{\varepsilon}$. 
Four in-vacuum rod electrodes (gray bars) generate dc electric fields along the $y$ axis.
The angle $\phi$ between $\bm{\varepsilon}$ and $\bm{E}$ can be used to adjust the first-order differential ac Stark shift between rotational states. 
A near-field dipole antenna emits 5.6 GHz microwaves (MWs) and couples the rotational states $\ket{J, m_J}$ (black lines) shown in (b). 
Blue boxes: nuclear spin states couple to rotation and mix in the $J=1$ manifold.
A dc Stark shift $\Delta$ splits the $\ket{1,0}$ and $\ket{1,\pm1}$ states.
}
\end{figure}

Our experiments begin with the preparation of ultracold $^{23}$Na$^{40}$K molecules in the rovibronic ground state at 300~nK \cite{Seesselberg2018} in several layers of a 1D lattice, see Fig. \ref{fig:schematic}(a).
The lattice is generated by a single, linearly polarized 1550~nm retro-reflected laser beam that propagates along the $z$ axis, which is also the direction of an 86~G magnetic field required for the molecule production. 
The polarization of the lattice beam can be adjusted with a half-wave plate within an uncertainty of $0.5$ degrees.
Initially, the molecules are prepared in the $\ket{J,m_J, m_{I, \mathrm{Na}}, m_{I, \mathrm{K}}} =\ket{0,0,-1/2,-4}$ hyperfine state which will be referred to as the ground state $\ket{\downarrow}$.
Here, $m_I$ denotes the projections of the nuclear spins $I_{\mathrm{Na}}=3/2$ \cite{Steck2003} and $I_{\mathrm{K}}=4$ \cite{Tiecke2010} onto the electric field axis.
A dc electric field along the $y$ axis is generated by applying voltages to four in-vacuum rod electrodes.
Eight additional auxiliary electrodes compensate residual electric field gradients to below 0.5~$\text{V/cm}^2$\cite{Supplement}.

As shown in Fig. \ref{fig:schematic}(b), molecules in the $J=0$  manifold can be coupled to the first excited rotational manifold $\ket{1, (0,\pm1)}$ via MW radiation with a frequency of $2 B_{\mathrm{rot}}/h\approx 5.6$~GHz \cite{Will2016}, $B_{\mathrm{rot}}$ denotes the rotational constant.
\begin{figure}
\includegraphics{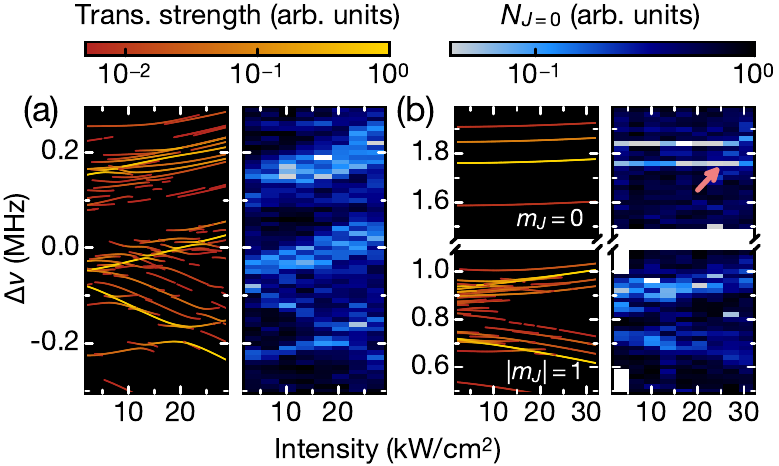}
\caption{\label{fig:acstark}
The ac Stark maps of the $J=0$ to $J=1$ transition manifold for two electric field strengths.
(a)
$E=$8.8~V/cm. 
Left panel: transition frequencies from the $\ket{\downarrow}$ state to the $J=1$ manifold as a function of light intensity.
The normalized transition strengths are encoded by line color. 
Only transitions stronger than 0.5\% are shown.
Right panel: molecule loss spectroscopy.
Molecules remaining in $\ket{\downarrow}$ after a MW sweep are recorded (blue).
(b) 
$E=$101.3 V/cm at magic trapping conditions. 
The $m_J=0$ component (upper panel) is separated from the $m_J = \pm1$ components (lower panel) by the dc Stark shift $\Delta$. 
Consequently, the hyperfine structure of the $m_J=0$ manifold is simplified to two strong lines.
Their transition frequency is almost independent of intensity.
The red arrow denotes the $\ket{\uparrow}$ state that will be used in the following.
Theoretical (experimental) data in both subfigures are normalized to the same maximum transition strength (detected atom number).
}
\end{figure}
There are $(2I_{\mathrm{Na}} + 1)(2I_{\mathrm{K}} +1) = 36$ hyperfine states in the $J=0$  manifold and 108 hyperfine states in the $J=1$ manifold.  
The nuclear spins in the $J=1$ manifold couple to rotation predominantly via the nuclear electric quadrupole moment. 
Furthermore, the trapping light field couples different $m_J$ states \cite{Gregory2017, Blackmore2018}. 
Subsequently, the hyperfine levels in the excited states are mixed and their energies show many avoided crossings as a function of light field intensity, see the left panel of Fig. \ref{fig:acstark}(a). 
Because of the strong mixing of the hyperfine levels, transition bands emerge rather than transition lines.
Even when the first-order differential light shift is canceled \cite{Kotochigova2010,Blackmore2018}, rotational states can therefore still rapidly dephase in an inhomogeneous optical trap. 
The right panel shows the result of the corresponding MW loss spectroscopy. 
In order to couple to states with different transition strengths, while maintaining good spectral resolution, we sweep the MW frequency across 10~kHz in 1.15~ms.
The Rabi frequency for the strongest transition is 4.0~kHz.
Whenever a reduction in $\ket{\downarrow}$ molecules is detected, a transition to $J = 1$ has occurred \cite{Supplement}.

In the presence of an electric field $E=101.3$~V/cm [see Fig. \ref{fig:acstark}(b)], the $m_J = 0$ states separate from the nearly degenerate $m_J=\pm1$ states due to the dc Stark splitting. 
Because this splitting is larger than all other interactions for electric fields as low as 60~V/cm, the nuclear spins decouple from the rotation, thus simplifying the ac Stark map. 
In addition, the rotation is decoupled from the light field, thereby reducing the curvature of the transition frequencies of these states.
Simultaneously, the polarization of the lattice beam is set to the magic angle with respect to the static field $\bm{E}$, thereby realizing a spin-decoupled magic trap.

In the following, we focus on $\ket{\uparrow}$, the hyperfine state of $J=1$, $m_J = 0$ with the largest transition strength. 
\begin{figure}
\includegraphics{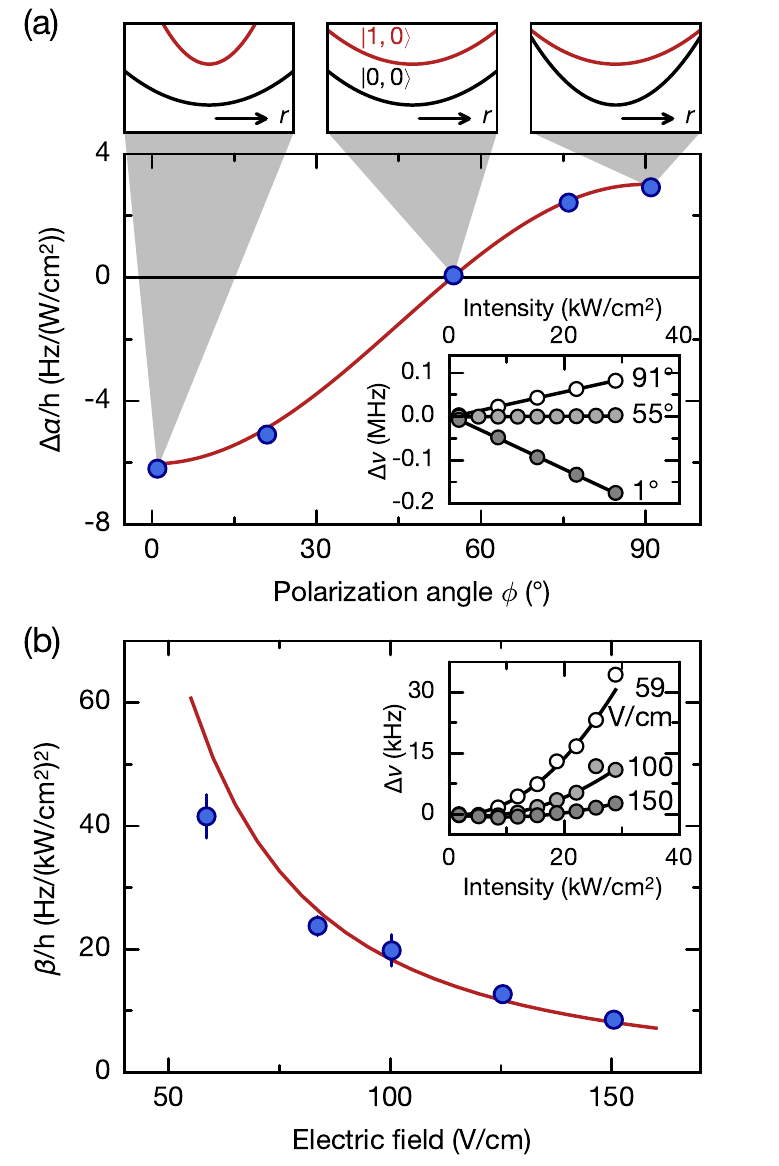}
\caption{\label{fig:magictrap}
Spin-decoupled magic trap.
Blue circles are measurements, red lines are theory, black lines are fits of the data to Eq. (\ref{eq:pol}).
(a) 
Differential polarizability $\Delta \alpha$ for various lattice polarization angles at $E=144.3$~V/cm. 
Inset: exemplary ac Stark data to extract $\Delta\alpha$. 
Top panels: Schematics of trapping potentials of $\ket{\downarrow}$ (black lines) or $\ket{\uparrow}$ (red lines), which depend on the polarization angle $\phi$. 
At approximately 54$^\circ$, a magic trapping condition is fulfilled.
(b)
Hyperpolarizability $\beta$ at the magic angle for various dc electric fields.
Inset: exemplary ac Stark data to extract $\beta$ for three electric fields.
All error bars were calculated from the covariance matrix of the fits.
}
\end{figure}
The dependence of the transition frequency $\nu$ on the light intensity $I$, the polarization angle $\phi$, and the electric field $E$ can be approximated by
\begin{equation}
\label{eq:pol}
\Delta \nu = \nu - \nu_{0} = \frac{1}{h}[\Delta \alpha(\phi) I + \beta(E,\phi) I^2  + \mathcal{O} (I^3)],
\end{equation}
where $\nu_{0}$ denotes the transition frequency at $I=0$, $\Delta \alpha = \alpha_{\ket{\downarrow}} - \alpha_{\ket{\uparrow}}$ is the differential polarizability, and $\beta$ is the hyperpolarizability of $J=1$ as $\beta \approx 0$ for $J=0$. 
Specifically, $\Delta\alpha(\phi) = 2/15 \times (1-3\cos^2\phi)\Delta\alpha_\mathrm{ele}$, where $\Delta\alpha_\mathrm{ele}= h\times22$~Hz/(W/cm$^2$) \cite{Neyenhuis2012,Supplement}.
To characterize the magic angle for this transition, we work at $E=144.3$~V/cm and use a $\pi$ pulse for the MW spectroscopy, see Fig. \ref{fig:magictrap}(a). 
For each $\phi$ we measure the transition frequency $\nu$ as a function of trap intensity and find differential polarizabilities that agree well with theory.
The magic condition $\Delta \alpha =0$ occurs for $\phi=54.0(5)^\circ$. 

To determine $\beta$, the same $\pi$-pulse spectroscopy, albeit with higher frequency resolution, is employed at $\phi = 54^\circ$, and for various electric fields, see the inset of Fig. \ref{fig:magictrap}(b). 
We extract $\beta$ (blue circles) by fitting Eq.~(\ref{eq:pol}) to our data and find that it decreases with increasing $E$. 
If $d_0 E \ll B_\mathrm{rot}$ and $d_0^2E^2/B_\mathrm{rot}$ is much larger than $\Delta\alpha_\mathrm{ele}I$ and the Zeeman splitting of $m_J$ states with the same hyperfine character at $E=0$, and away from any spectral crossings,  $\beta$ can be derived from second-order perturbation of the energy as
\begin{equation}
\beta(E,\phi) = \frac{4}{15}\sin^2{(2\phi)}\frac{\Delta\alpha_\mathrm{ele}^2B_\mathrm{rot}}{d_0^2E^2} \;,
\label{eq:beta}
\end{equation}
shown as red line in Fig. \ref{fig:magictrap}(b) for our parameters.

\begin{figure*}
\includegraphics{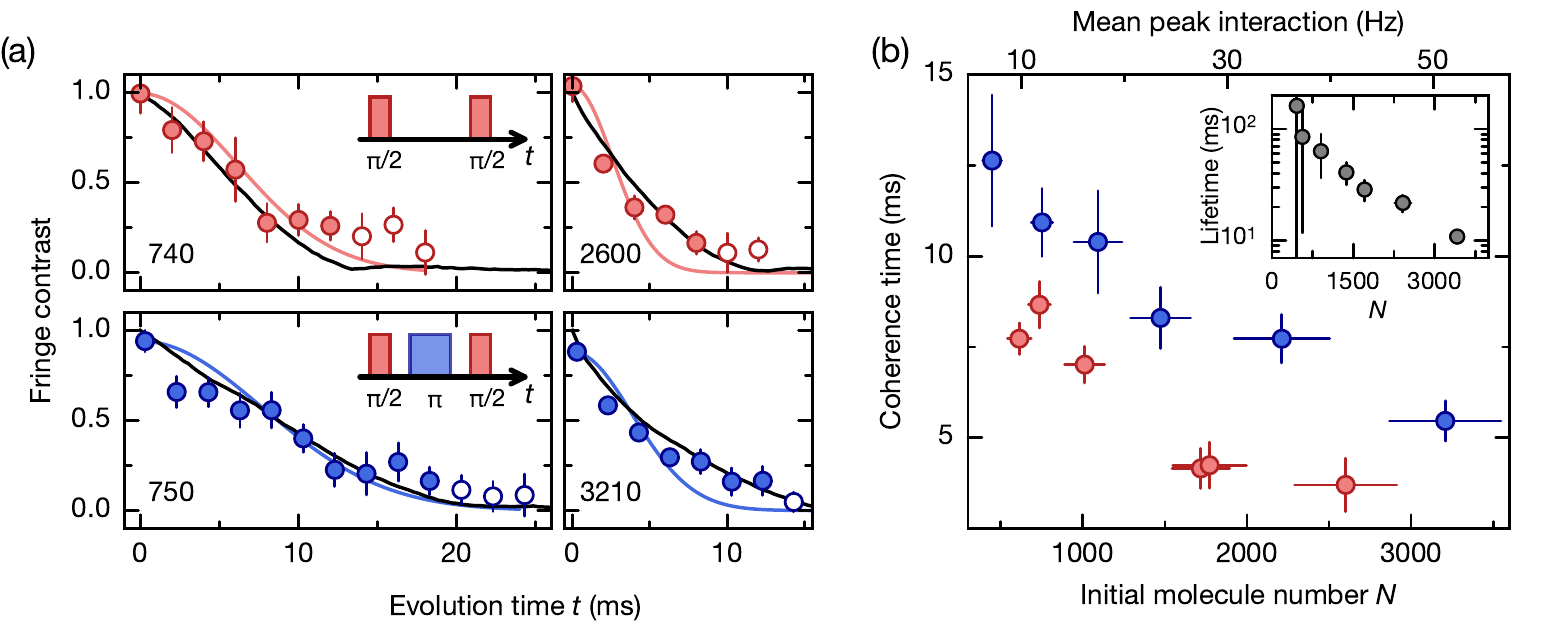}
\caption{\label{fig:coherence}
Rotational coherence in the spin-decoupled magic trap.
(a) 
Contrast of the Ramsey (red circles) and spin-echo (blue circles) fringes for various evolution times $t$.
MW pulse sequences are shown in the upper right, initial molecular numbers in the lower left corner.
Data points below the  bias cutoff (empty circles) are excluded for extracting the coherence time with a Gaussian fit (colored lines).  
MACE simulations (black lines) of a dipolar Hamiltonian with a dephasing rate of $h \times 21(2)$~Hz ($h \times 35(2)$~Hz) describe our observations for the Ramsey (spin-echo) experiments well. 
(b)
Ramsey (red) and spin-echo (blue) coherence times at various molecular densities.
The mean dipolar interaction strength at the center of the cloud is indicated on the secondary $x$ axis \cite{Supplement}.
The coherence time is not limited by the $1/e$ lifetime of the rotational superposition, shown in the inset.
All error bars are calculated from the covariance matrix of the fits and denote one standard deviation. 
} 
\end{figure*}

Next, we study the rotational coherence in the spin-decoupled magic trap, see Fig. \ref{fig:coherence}.
We use Ramsey and spin-echo pulse sequences \cite{Ramsey1950} and work at $I=3.4$~kW/cm$^2$.
We set $E=68.3$~V/cm, which is large enough to decouple the $\ket{\uparrow}$ state and small enough to minimize inhomogeneous broadening or temporal noise of the dc Stark shift. 
The MW frequency $\nu$ is set to resonance. 
We scan the relative phase $\Delta \theta$ between the first and second $\pi/2$ pulse at a fixed evolution time $t$ to obtain Ramsey interference fringes.
Each fringe is described by 
\begin{equation}
N_{\ket{\downarrow}}(\Delta\theta, t)=\frac{N_{\mathrm{tot}}(t)}{2}[1-c(t)\cos(\Delta\theta+\theta_0)],
\end{equation}
where $c(t)$ is the measured contrast, $N_{\mathrm{tot}}=N_{\ket{\downarrow}}+N_{\ket{\uparrow}}$ is the total molecule number and $\theta_0$ is a phase offset due to small detunings of the MW, e.\ g.\ due to electric field changes. 
We measure $c(t)$ for various molecule numbers, see Fig. \ref{fig:coherence}(a), and fit a Gaussian to extract the coherence time.
Because $c$ is strictly positive in the fringe fitting, it biases the coherence time when the fringe amplitude becomes comparable to molecule number fluctuations.
We therefore estimate the bias $\Delta c$ for each data point individually and exclude data taken after the first point where $c < 1.5\Delta c$ \cite{Supplement}.
The Ramsey coherence time $\tau_c$, here defined as the $1/e$ time of the fit, amounts to 8.7(6)~ms for a low molecule number $N_{\mathrm{tot}}=740(70)$, which is six times larger than previously achieved coherence times \cite{Neyenhuis2012,Blackmore2018}. 

Residual single particle dephasing could arise due to residual differential light shifts, electric field gradients, and shot-to-shot fluctuations of the electric field.  
By adding a $\pi$ pulse in the middle of the evolution, we obtain a spin-echo sequence that cancels the slowly varying contributions to the single particle dephasing and allows us to increase the coherence time to $\tau_c = 13(2)$ ms for low initial molecule numbers.
Note that the molecules in this work are moving with the trapping period of $T_{\mathrm{trap}}=16$~ms in the horizontal planes, which are weakly confined by the 1D lattice. 
Spin echo fails to suppress or even enhances the single particle dephasing when the evolution time is close to the trapping period \cite{Koller2015}.
This explains why the maximum coherence time observed in our experiment remains below $T_{\mathrm{trap}}$.

Furthermore, we find that the coherence time depends on the initial molecule number and thus on density, see Fig.~\ref{fig:coherence}(b) \cite{Supplement}. 
There could be several reasons for this, for example a loss of molecules. 
We measure an intrastate inelastic collision rate of below 3 Hz, because these collisions are suppressed by the $p$-wave barrier.
Thus, interstate inelastic collisions dominate, which leads to equal loss of $\ket{\downarrow}$ and $\ket{\uparrow}$ and does not reduce the fringe contrast. 
Furthermore, this two-body loss occurs on much longer time scales than the decoherence, see inset of Fig. \ref{fig:coherence}(b). 
Another reason for the density dependent decoherence is the strong dipolar interaction present in the system. 

To qualitatively understand the decoherence of the molecular rotation caused by dipolar interactions, we use the moving average cluster expansion (MACE) method \cite{Hazzard2014a} to simulate the spin dynamics of randomly distributed molecules in bulk during the Ramsey or spin-echo interferometry \cite{Supplement}.
Neglecting loss and molecular motion, the system can be described by the following Hamiltonian
\begin{eqnarray}
H=\sum_{i>j}\frac{U_{ij}}{2}(\hat{S}^+_i\hat{S}^-_j+\mathrm{h.c.})+\sum_{i}\Delta(\mathbfit{r}_i)\hat{S}^z_i,
\label{eqs1}
\end{eqnarray}
where the first term describes the dipolar spin-exchange interaction, where $\hat{S}^{\pm}_i$ and $\hat{S}^z_i$ are the spin-1/2 angular momentum operators of molecule $i$ in position $\mathbfit{r}_i$, $U_{ij}=2 d^2_ {\uparrow\downarrow}/{(4\pi\varepsilon_0)}\times{(1-3\cos^2\Theta_{ij})}/{(|\mathbfit{r}_i-\mathbfit{r}_j|^3)}$ is the dipole-dipole interaction strength between molecules $i$ and $j$, $d_{\uparrow\downarrow}=\sqrt{1/3}d_0$ is the transition dipole moment between $\ket{\downarrow}$ and $\ket{\uparrow}$ \cite{Micheli2007,Gorshkov2011b}, $\varepsilon_0$ is the vacuum permittivity, and $\Theta_{ij}$ is the angle between the vector connecting molecules $i$ and $j$ and the quantization axis.
The second term describes the coupling to external fields, where $\Delta(\mathbfit{r}_i)$ is a spatially dependent detuning of the microwave transition \cite{Supplement}.
We use $\Delta(\mathbfit{r}_i)$ to emulate the effects of single particle dephasing, especially the uncanceled, movement-induced, time-dependent gradient in the spin-echo case.
By modeling this inhomogeneity as an effective external field gradient,
the Ramsey (spin-echo) signal with very low molecule number, for which the dipolar interactions can be neglected, can be reproduced.
The corresponding single particle dephasing rate is $h\times35(2)$~Hz ($h\times21(2)$~Hz), which corresponds to a dephasing time of 9~(15)~ms for the Ramsey (spin-echo) case.
Using these values as input for the MACE model leads to simulation results that are consistent with experimental observations for all other densities [black lines, see \cite{Supplement} for all data sets], four of which are shown in Fig. \ref{fig:coherence}(a). 
This indicates that dipolar interactions are the dominant source of the density-dependent decoherence.
A theoretical model tailored to the trap geometry discussed in this work could improve the understanding of how molecular loss, motion and contact interaction modify the spin dynamics in a bulk gas of polar molecules.

In conclusion, we presented a novel trapping technique for rotating molecules that cancels differential polarizability and reduces the hyperpolarizability. 
With this method, applicable to a broad range of polar molecules, a density dependence of the  rotational coherence time is observed, which is attributed to molecular dipole-dipole interactions and characterized using a simple numerical model.
For low density, coherence times as large as 13(2)~ms were obtained in the molecular clouds. 
This opens up exciting possibilities for further experiments.
The interplay between the kinetic energy and dipolar interaction could be studied in a bulk gas of molecules.
If even longer coherence times are required, a spin-decoupled magic 3D optical lattice could be used to freeze out any molecular motion.
This seems very promising because rotational coherence times of about 100 ms were already achieved in a non-spin-decoupled  magic 3D lattice \cite{Yan2013a}.
For a near unity filling \NaK gas in a 3D optical lattice, we expect a dipolar interaction energy on the order of $h \times 1$~kHz, much stronger than the single particle dephasing. 
This will allow the observation of new states of dipolar quantum matter, e.g. a condensate of rotational excitations \cite{Kwasigroch2014}.

\begin{acknowledgements}
We thank Kaden Hazzard, Tao Shi, Gang-Qin Liu and Richard Schmidt for inspiring discussions, Jun Ye for help with the high-precision high-voltage source, 
Marcel Duda and Scott Eustice for careful reading of the manuscript, and Nikolaus Buchheim and Zhen-Kai Lu for their contributions to the experimental setup. 
The MPQ team gratefully acknowledges support from the DFG (FOR 2247) and the EU (UQUAM). 
Work at Temple University is supported by the ARO-MURI Grant No. W911NF-14-1-0378, the ARO Grant No. W911NF-17-1-0563, the AFOSR Grant No. FA9550-14-1-0321, and the NSF Grant No.PHY-1619788.\\
\end{acknowledgements}

\nocite{*}
\bibliography{PaperNo2_clean}

\begin{thebibliography}{45}%
\makeatletter
\providecommand \@ifxundefined [1]{%
 \@ifx{#1\undefined}
}%
\providecommand \@ifnum [1]{%
 \ifnum #1\expandafter \@firstoftwo
 \else \expandafter \@secondoftwo
 \fi
}%
\providecommand \@ifx [1]{%
 \ifx #1\expandafter \@firstoftwo
 \else \expandafter \@secondoftwo
 \fi
}%
\providecommand \natexlab [1]{#1}%
\providecommand \enquote  [1]{``#1''}%
\providecommand \bibnamefont  [1]{#1}%
\providecommand \bibfnamefont [1]{#1}%
\providecommand \citenamefont [1]{#1}%
\providecommand \href@noop [0]{\@secondoftwo}%
\providecommand \href [0]{\begingroup \@sanitize@url \@href}%
\providecommand \@href[1]{\@@startlink{#1}\@@href}%
\providecommand \@@href[1]{\endgroup#1\@@endlink}%
\providecommand \@sanitize@url [0]{\catcode `\\12\catcode `\$12\catcode
  `\&12\catcode `\#12\catcode `\^12\catcode `\_12\catcode `\%12\relax}%
\providecommand \@@startlink[1]{}%
\providecommand \@@endlink[0]{}%
\providecommand \url  [0]{\begingroup\@sanitize@url \@url }%
\providecommand \@url [1]{\endgroup\@href {#1}{\urlprefix }}%
\providecommand \urlprefix  [0]{URL }%
\providecommand \Eprint [0]{\href }%
\providecommand \doibase [0]{http://dx.doi.org/}%
\providecommand \selectlanguage [0]{\@gobble}%
\providecommand \bibinfo  [0]{\@secondoftwo}%
\providecommand \bibfield  [0]{\@secondoftwo}%
\providecommand \translation [1]{[#1]}%
\providecommand \BibitemOpen [0]{}%
\providecommand \bibitemStop [0]{}%
\providecommand \bibitemNoStop [0]{.\EOS\space}%
\providecommand \EOS [0]{\spacefactor3000\relax}%
\providecommand \BibitemShut  [1]{\csname bibitem#1\endcsname}%
\let\auto@bib@innerbib\@empty
\bibitem [{\citenamefont {Ni}\ \emph {et~al.}(2008)\citenamefont {Ni},
  \citenamefont {Ospelkaus}, \citenamefont {De~Miranda}, \citenamefont {Pe'er},
  \citenamefont {Neyenhuis}, \citenamefont {Zirbel}, \citenamefont
  {Kotochigova}, \citenamefont {Julienne}, \citenamefont {Jin},\ and\
  \citenamefont {Ye}}]{Ni2008}%
  \BibitemOpen
  \bibfield  {author} {\bibinfo {author} {\bibfnamefont {K.-K.}\ \bibnamefont
  {Ni}}, \bibinfo {author} {\bibfnamefont {S.}~\bibnamefont {Ospelkaus}},
  \bibinfo {author} {\bibfnamefont {M.}~\bibnamefont {De~Miranda}}, \bibinfo
  {author} {\bibfnamefont {A.}~\bibnamefont {Pe'er}}, \bibinfo {author}
  {\bibfnamefont {B.}~\bibnamefont {Neyenhuis}}, \bibinfo {author}
  {\bibfnamefont {J.}~\bibnamefont {Zirbel}}, \bibinfo {author} {\bibfnamefont
  {S.}~\bibnamefont {Kotochigova}}, \bibinfo {author} {\bibfnamefont
  {P.}~\bibnamefont {Julienne}}, \bibinfo {author} {\bibfnamefont
  {D.}~\bibnamefont {Jin}}, \ and\ \bibinfo {author} {\bibfnamefont
  {J.}~\bibnamefont {Ye}},\ }\href {\doibase 10.1126/science.1163861}
  {\bibfield  {journal} {\bibinfo  {journal} {Science}\ }\textbf {\bibinfo
  {volume} {322}},\ \bibinfo {pages} {231} (\bibinfo {year}
  {2008})}\BibitemShut {NoStop}%
\bibitem [{\citenamefont {Shuman}\ \emph {et~al.}(2010)\citenamefont {Shuman},
  \citenamefont {Barry},\ and\ \citenamefont {DeMille}}]{Shuman2010}%
  \BibitemOpen
  \bibfield  {author} {\bibinfo {author} {\bibfnamefont {E.~S.}\ \bibnamefont
  {Shuman}}, \bibinfo {author} {\bibfnamefont {J.~F.}\ \bibnamefont {Barry}}, \
  and\ \bibinfo {author} {\bibfnamefont {D.}~\bibnamefont {DeMille}},\ }\href
  {\doibase 10.1038/nature09443} {\bibfield  {journal} {\bibinfo  {journal}
  {Nature}\ }\textbf {\bibinfo {volume} {467}},\ \bibinfo {pages} {820}
  (\bibinfo {year} {2010})}\BibitemShut {NoStop}%
\bibitem [{\citenamefont {Takekoshi}\ \emph {et~al.}(2014)\citenamefont
  {Takekoshi}, \citenamefont {Reichs\"ollner}, \citenamefont {Schindewolf},
  \citenamefont {Hutson}, \citenamefont {Le~Sueur}, \citenamefont {Dulieu},
  \citenamefont {Ferlaino}, \citenamefont {Grimm},\ and\ \citenamefont
  {N\"agerl}}]{Takekoshi2014}%
  \BibitemOpen
  \bibfield  {author} {\bibinfo {author} {\bibfnamefont {T.}~\bibnamefont
  {Takekoshi}}, \bibinfo {author} {\bibfnamefont {L.}~\bibnamefont
  {Reichs\"ollner}}, \bibinfo {author} {\bibfnamefont {A.}~\bibnamefont
  {Schindewolf}}, \bibinfo {author} {\bibfnamefont {J.~M.}\ \bibnamefont
  {Hutson}}, \bibinfo {author} {\bibfnamefont {C.~R.}\ \bibnamefont
  {Le~Sueur}}, \bibinfo {author} {\bibfnamefont {O.}~\bibnamefont {Dulieu}},
  \bibinfo {author} {\bibfnamefont {F.}~\bibnamefont {Ferlaino}}, \bibinfo
  {author} {\bibfnamefont {R.}~\bibnamefont {Grimm}}, \ and\ \bibinfo {author}
  {\bibfnamefont {H.-C.}\ \bibnamefont {N\"agerl}},\ }\href {\doibase
  10.1103/PhysRevLett.113.205301} {\bibfield  {journal} {\bibinfo  {journal}
  {Phys. Rev. Lett.}\ }\textbf {\bibinfo {volume} {113}},\ \bibinfo {pages}
  {205301} (\bibinfo {year} {2014})}\BibitemShut {NoStop}%
\bibitem [{\citenamefont {Molony}\ \emph {et~al.}(2014)\citenamefont {Molony},
  \citenamefont {Gregory}, \citenamefont {Ji}, \citenamefont {Lu},
  \citenamefont {K{\"{o}}ppinger}, \citenamefont {{Le Sueur}}, \citenamefont
  {Blackley}, \citenamefont {Hutson},\ and\ \citenamefont
  {Cornish}}]{Molony2014}%
  \BibitemOpen
  \bibfield  {author} {\bibinfo {author} {\bibfnamefont {P.~K.}\ \bibnamefont
  {Molony}}, \bibinfo {author} {\bibfnamefont {P.~D.}\ \bibnamefont {Gregory}},
  \bibinfo {author} {\bibfnamefont {Z.}~\bibnamefont {Ji}}, \bibinfo {author}
  {\bibfnamefont {B.}~\bibnamefont {Lu}}, \bibinfo {author} {\bibfnamefont
  {M.~P.}\ \bibnamefont {K{\"{o}}ppinger}}, \bibinfo {author} {\bibfnamefont
  {C.~R.}\ \bibnamefont {{Le Sueur}}}, \bibinfo {author} {\bibfnamefont
  {C.~L.}\ \bibnamefont {Blackley}}, \bibinfo {author} {\bibfnamefont {J.~M.}\
  \bibnamefont {Hutson}}, \ and\ \bibinfo {author} {\bibfnamefont {S.~L.}\
  \bibnamefont {Cornish}},\ }\href {\doibase 10.1103/PhysRevLett.113.255301}
  {\bibfield  {journal} {\bibinfo  {journal} {Phys. Rev. Lett.}\ }\textbf
  {\bibinfo {volume} {113}},\ \bibinfo {pages} {255301} (\bibinfo {year}
  {2014})}\BibitemShut {NoStop}%
\bibitem [{\citenamefont {Park}\ \emph {et~al.}(2015)\citenamefont {Park},
  \citenamefont {Will},\ and\ \citenamefont {Zwierlein}}]{Park2015}%
  \BibitemOpen
  \bibfield  {author} {\bibinfo {author} {\bibfnamefont {J.~W.}\ \bibnamefont
  {Park}}, \bibinfo {author} {\bibfnamefont {S.~A.}\ \bibnamefont {Will}}, \
  and\ \bibinfo {author} {\bibfnamefont {M.~W.}\ \bibnamefont {Zwierlein}},\
  }\href {\doibase 10.1103/PhysRevLett.114.205302} {\bibfield  {journal}
  {\bibinfo  {journal} {Phys. Rev. Lett.}\ }\textbf {\bibinfo {volume} {114}},\
  \bibinfo {pages} {205302} (\bibinfo {year} {2015})}\BibitemShut {NoStop}%
\bibitem [{\citenamefont {Guo}\ \emph {et~al.}(2016)\citenamefont {Guo},
  \citenamefont {Zhu}, \citenamefont {Lu}, \citenamefont {Ye}, \citenamefont
  {Wang}, \citenamefont {Vexiau}, \citenamefont {Bouloufa-Maafa}, \citenamefont
  {Qu\'em\'ener}, \citenamefont {Dulieu},\ and\ \citenamefont
  {Wang}}]{Guo2016}%
  \BibitemOpen
  \bibfield  {author} {\bibinfo {author} {\bibfnamefont {M.}~\bibnamefont
  {Guo}}, \bibinfo {author} {\bibfnamefont {B.}~\bibnamefont {Zhu}}, \bibinfo
  {author} {\bibfnamefont {B.}~\bibnamefont {Lu}}, \bibinfo {author}
  {\bibfnamefont {X.}~\bibnamefont {Ye}}, \bibinfo {author} {\bibfnamefont
  {F.}~\bibnamefont {Wang}}, \bibinfo {author} {\bibfnamefont {R.}~\bibnamefont
  {Vexiau}}, \bibinfo {author} {\bibfnamefont {N.}~\bibnamefont
  {Bouloufa-Maafa}}, \bibinfo {author} {\bibfnamefont {G.}~\bibnamefont
  {Qu\'em\'ener}}, \bibinfo {author} {\bibfnamefont {O.}~\bibnamefont
  {Dulieu}}, \ and\ \bibinfo {author} {\bibfnamefont {D.}~\bibnamefont
  {Wang}},\ }\href {\doibase 10.1103/PhysRevLett.116.205303} {\bibfield
  {journal} {\bibinfo  {journal} {Phys. Rev. Lett.}\ }\textbf {\bibinfo
  {volume} {116}},\ \bibinfo {pages} {205303} (\bibinfo {year}
  {2016})}\BibitemShut {NoStop}%
\bibitem [{\citenamefont {Prehn}\ \emph {et~al.}(2016)\citenamefont {Prehn},
  \citenamefont {Ibr{\"{u}}gger}, \citenamefont {Gl{\"{o}}ckner}, \citenamefont
  {Rempe},\ and\ \citenamefont {Zeppenfeld}}]{Prehn2016}%
  \BibitemOpen
  \bibfield  {author} {\bibinfo {author} {\bibfnamefont {A.}~\bibnamefont
  {Prehn}}, \bibinfo {author} {\bibfnamefont {M.}~\bibnamefont
  {Ibr{\"{u}}gger}}, \bibinfo {author} {\bibfnamefont {R.}~\bibnamefont
  {Gl{\"{o}}ckner}}, \bibinfo {author} {\bibfnamefont {G.}~\bibnamefont
  {Rempe}}, \ and\ \bibinfo {author} {\bibfnamefont {M.}~\bibnamefont
  {Zeppenfeld}},\ }\href {\doibase 10.1103/PhysRevLett.116.063005} {\bibfield
  {journal} {\bibinfo  {journal} {Phys. Rev. Lett.}\ }\textbf {\bibinfo
  {volume} {116}},\ \bibinfo {pages} {063005} (\bibinfo {year}
  {2016})}\BibitemShut {NoStop}%
\bibitem [{\citenamefont {Rvachov}\ \emph {et~al.}(2017)\citenamefont
  {Rvachov}, \citenamefont {Son}, \citenamefont {Sommer}, \citenamefont
  {Ebadi}, \citenamefont {Park}, \citenamefont {Zwierlein}, \citenamefont
  {Ketterle},\ and\ \citenamefont {Jamison}}]{Rvachov2017}%
  \BibitemOpen
  \bibfield  {author} {\bibinfo {author} {\bibfnamefont {T.~M.}\ \bibnamefont
  {Rvachov}}, \bibinfo {author} {\bibfnamefont {H.}~\bibnamefont {Son}},
  \bibinfo {author} {\bibfnamefont {A.~T.}\ \bibnamefont {Sommer}}, \bibinfo
  {author} {\bibfnamefont {S.}~\bibnamefont {Ebadi}}, \bibinfo {author}
  {\bibfnamefont {J.~J.}\ \bibnamefont {Park}}, \bibinfo {author}
  {\bibfnamefont {M.~W.}\ \bibnamefont {Zwierlein}}, \bibinfo {author}
  {\bibfnamefont {W.}~\bibnamefont {Ketterle}}, \ and\ \bibinfo {author}
  {\bibfnamefont {A.~O.}\ \bibnamefont {Jamison}},\ }\href {\doibase
  10.1103/PhysRevLett.119.143001} {\bibfield  {journal} {\bibinfo  {journal}
  {Phys. Rev. Lett.}\ }\textbf {\bibinfo {volume} {119}},\ \bibinfo {pages}
  {143001} (\bibinfo {year} {2017})}\BibitemShut {NoStop}%
\bibitem [{\citenamefont {See{\ss}elberg}\ \emph {et~al.}(2018)\citenamefont
  {See{\ss}elberg}, \citenamefont {Buchheim}, \citenamefont {Lu}, \citenamefont
  {Schneider}, \citenamefont {Luo}, \citenamefont {Tiemann}, \citenamefont
  {Bloch},\ and\ \citenamefont {Gohle}}]{Seesselberg2018}%
  \BibitemOpen
  \bibfield  {author} {\bibinfo {author} {\bibfnamefont {F.}~\bibnamefont
  {See{\ss}elberg}}, \bibinfo {author} {\bibfnamefont {N.}~\bibnamefont
  {Buchheim}}, \bibinfo {author} {\bibfnamefont {Z.-K.}\ \bibnamefont {Lu}},
  \bibinfo {author} {\bibfnamefont {T.}~\bibnamefont {Schneider}}, \bibinfo
  {author} {\bibfnamefont {X.-Y.}\ \bibnamefont {Luo}}, \bibinfo {author}
  {\bibfnamefont {E.}~\bibnamefont {Tiemann}}, \bibinfo {author} {\bibfnamefont
  {I.}~\bibnamefont {Bloch}}, \ and\ \bibinfo {author} {\bibfnamefont
  {C.}~\bibnamefont {Gohle}},\ }\href {\doibase 10.1103/PhysRevA.97.013405}
  {\bibfield  {journal} {\bibinfo  {journal} {Phys. Rev. A}\ }\textbf {\bibinfo
  {volume} {97}},\ \bibinfo {pages} {013405} (\bibinfo {year}
  {2018})}\BibitemShut {NoStop}%
\bibitem [{\citenamefont {Anderegg}\ \emph {et~al.}(2018)\citenamefont
  {Anderegg}, \citenamefont {Augenbraun}, \citenamefont {Bao}, \citenamefont
  {Burchesky}, \citenamefont {Cheuk}, \citenamefont {Ketterle},\ and\
  \citenamefont {Doyle}}]{Anderegg2018}%
  \BibitemOpen
  \bibfield  {author} {\bibinfo {author} {\bibfnamefont {L.}~\bibnamefont
  {Anderegg}}, \bibinfo {author} {\bibfnamefont {B.~L.}\ \bibnamefont
  {Augenbraun}}, \bibinfo {author} {\bibfnamefont {Y.}~\bibnamefont {Bao}},
  \bibinfo {author} {\bibfnamefont {S.}~\bibnamefont {Burchesky}}, \bibinfo
  {author} {\bibfnamefont {L.~W.}\ \bibnamefont {Cheuk}}, \bibinfo {author}
  {\bibfnamefont {W.}~\bibnamefont {Ketterle}}, \ and\ \bibinfo {author}
  {\bibfnamefont {J.~M.}\ \bibnamefont {Doyle}},\ }\href {\doibase
  10.1038/s41567-018-0191-z} {\bibfield  {journal} {\bibinfo  {journal} {Nat.
  Phys.}\ }\textbf {\bibinfo {volume} {14}},\ \bibinfo {pages} {890} (\bibinfo
  {year} {2018})}\BibitemShut {NoStop}%
\bibitem [{\citenamefont {Collopy}\ \emph {et~al.}(2018)\citenamefont
  {Collopy}, \citenamefont {Ding}, \citenamefont {Wu}, \citenamefont
  {Finneran}, \citenamefont {Anderegg}, \citenamefont {Augenbraun},
  \citenamefont {Doyle},\ and\ \citenamefont {Ye}}]{Collopy2018}%
  \BibitemOpen
  \bibfield  {author} {\bibinfo {author} {\bibfnamefont {A.~L.}\ \bibnamefont
  {Collopy}}, \bibinfo {author} {\bibfnamefont {S.}~\bibnamefont {Ding}},
  \bibinfo {author} {\bibfnamefont {Y.}~\bibnamefont {Wu}}, \bibinfo {author}
  {\bibfnamefont {I.~A.}\ \bibnamefont {Finneran}}, \bibinfo {author}
  {\bibfnamefont {L.}~\bibnamefont {Anderegg}}, \bibinfo {author}
  {\bibfnamefont {B.~L.}\ \bibnamefont {Augenbraun}}, \bibinfo {author}
  {\bibfnamefont {J.~M.}\ \bibnamefont {Doyle}}, \ and\ \bibinfo {author}
  {\bibfnamefont {J.}~\bibnamefont {Ye}},\ }\href {\doibase
  10.1103/PhysRevLett.121.213201} {\bibfield  {journal} {\bibinfo  {journal}
  {Phys. Rev. Lett.}\ }\textbf {\bibinfo {volume} {121}},\ \bibinfo {pages}
  {213201} (\bibinfo {year} {2018})}\BibitemShut {NoStop}%
\bibitem [{\citenamefont {Moses}\ \emph {et~al.}(2017)\citenamefont {Moses},
  \citenamefont {Covey}, \citenamefont {Miecnikowski}, \citenamefont {Jin},\
  and\ \citenamefont {Ye}}]{Moses2016a}%
  \BibitemOpen
  \bibfield  {author} {\bibinfo {author} {\bibfnamefont {S.~A.}\ \bibnamefont
  {Moses}}, \bibinfo {author} {\bibfnamefont {J.~P.}\ \bibnamefont {Covey}},
  \bibinfo {author} {\bibfnamefont {M.~T.}\ \bibnamefont {Miecnikowski}},
  \bibinfo {author} {\bibfnamefont {D.~S.}\ \bibnamefont {Jin}}, \ and\
  \bibinfo {author} {\bibfnamefont {J.}~\bibnamefont {Ye}},\ }\href {\doibase
  10.1038/nphys3985} {\bibfield  {journal} {\bibinfo  {journal} {Nat. Phys.}\
  }\textbf {\bibinfo {volume} {13}},\ \bibinfo {pages} {13} (\bibinfo {year}
  {2017})}\BibitemShut {NoStop}%
\bibitem [{\citenamefont {{De Marco}}\ \emph {et~al.}(2018)\citenamefont {{De
  Marco}}, \citenamefont {Valtolina}, \citenamefont {Matsuda}, \citenamefont
  {Tobias}, \citenamefont {Covey},\ and\ \citenamefont {Ye}}]{DeMarco2018}%
  \BibitemOpen
  \bibfield  {author} {\bibinfo {author} {\bibfnamefont {L.}~\bibnamefont {{De
  Marco}}}, \bibinfo {author} {\bibfnamefont {G.}~\bibnamefont {Valtolina}},
  \bibinfo {author} {\bibfnamefont {K.}~\bibnamefont {Matsuda}}, \bibinfo
  {author} {\bibfnamefont {W.~G.}\ \bibnamefont {Tobias}}, \bibinfo {author}
  {\bibfnamefont {J.~P.}\ \bibnamefont {Covey}}, \ and\ \bibinfo {author}
  {\bibfnamefont {J.}~\bibnamefont {Ye}},\ }\href
  {http://arxiv.org/abs/1808.00028} {\bibfield  {journal} {\bibinfo  {journal}
  {arXiv:1808.00028}\ } (\bibinfo {year} {2018})}\BibitemShut {NoStop}%
\bibitem [{\citenamefont {Chotia}\ \emph {et~al.}(2012)\citenamefont {Chotia},
  \citenamefont {Neyenhuis}, \citenamefont {Moses}, \citenamefont {Yan},
  \citenamefont {Covey}, \citenamefont {Foss-Feig}, \citenamefont {Rey},
  \citenamefont {Jin},\ and\ \citenamefont {Ye}}]{Chotia2012}%
  \BibitemOpen
  \bibfield  {author} {\bibinfo {author} {\bibfnamefont {A.}~\bibnamefont
  {Chotia}}, \bibinfo {author} {\bibfnamefont {B.}~\bibnamefont {Neyenhuis}},
  \bibinfo {author} {\bibfnamefont {S.~A.}\ \bibnamefont {Moses}}, \bibinfo
  {author} {\bibfnamefont {B.}~\bibnamefont {Yan}}, \bibinfo {author}
  {\bibfnamefont {J.~P.}\ \bibnamefont {Covey}}, \bibinfo {author}
  {\bibfnamefont {M.}~\bibnamefont {Foss-Feig}}, \bibinfo {author}
  {\bibfnamefont {A.~M.}\ \bibnamefont {Rey}}, \bibinfo {author} {\bibfnamefont
  {D.~S.}\ \bibnamefont {Jin}}, \ and\ \bibinfo {author} {\bibfnamefont
  {J.}~\bibnamefont {Ye}},\ }\href {\doibase 10.1103/PhysRevLett.108.080405}
  {\bibfield  {journal} {\bibinfo  {journal} {Phys. Rev. Lett.}\ }\textbf
  {\bibinfo {volume} {108}},\ \bibinfo {pages} {080405} (\bibinfo {year}
  {2012})}\BibitemShut {NoStop}%
\bibitem [{\citenamefont {Yan}\ \emph {et~al.}(2013)\citenamefont {Yan},
  \citenamefont {Moses}, \citenamefont {Gadway}, \citenamefont {Covey},
  \citenamefont {Hazzard}, \citenamefont {Rey}, \citenamefont {Jin},\ and\
  \citenamefont {Ye}}]{Yan2013a}%
  \BibitemOpen
  \bibfield  {author} {\bibinfo {author} {\bibfnamefont {B.}~\bibnamefont
  {Yan}}, \bibinfo {author} {\bibfnamefont {S.~A.}\ \bibnamefont {Moses}},
  \bibinfo {author} {\bibfnamefont {B.}~\bibnamefont {Gadway}}, \bibinfo
  {author} {\bibfnamefont {J.~P.}\ \bibnamefont {Covey}}, \bibinfo {author}
  {\bibfnamefont {K.~R.~A.}\ \bibnamefont {Hazzard}}, \bibinfo {author}
  {\bibfnamefont {A.~M.}\ \bibnamefont {Rey}}, \bibinfo {author} {\bibfnamefont
  {D.~S.}\ \bibnamefont {Jin}}, \ and\ \bibinfo {author} {\bibfnamefont
  {J.}~\bibnamefont {Ye}},\ }\href {\doibase 10.1038/nature12483} {\bibfield
  {journal} {\bibinfo  {journal} {Nature}\ }\textbf {\bibinfo {volume} {501}},\
  \bibinfo {pages} {521} (\bibinfo {year} {2013})}\BibitemShut {NoStop}%
\bibitem [{\citenamefont {Ospelkaus}\ \emph {et~al.}(2010)\citenamefont
  {Ospelkaus}, \citenamefont {Ni}, \citenamefont {Qu{\'{e}}m{\'{e}}ner},
  \citenamefont {Neyenhuis}, \citenamefont {Wang}, \citenamefont {de~Miranda},
  \citenamefont {Bohn}, \citenamefont {Ye},\ and\ \citenamefont
  {Jin}}]{Ospelkaus2010}%
  \BibitemOpen
  \bibfield  {author} {\bibinfo {author} {\bibfnamefont {S.}~\bibnamefont
  {Ospelkaus}}, \bibinfo {author} {\bibfnamefont {K.-K.}\ \bibnamefont {Ni}},
  \bibinfo {author} {\bibfnamefont {G.}~\bibnamefont {Qu{\'{e}}m{\'{e}}ner}},
  \bibinfo {author} {\bibfnamefont {B.}~\bibnamefont {Neyenhuis}}, \bibinfo
  {author} {\bibfnamefont {D.}~\bibnamefont {Wang}}, \bibinfo {author}
  {\bibfnamefont {M.~H.~G.}\ \bibnamefont {de~Miranda}}, \bibinfo {author}
  {\bibfnamefont {J.~L.}\ \bibnamefont {Bohn}}, \bibinfo {author}
  {\bibfnamefont {J.}~\bibnamefont {Ye}}, \ and\ \bibinfo {author}
  {\bibfnamefont {D.~S.}\ \bibnamefont {Jin}},\ }\href {\doibase
  10.1103/PhysRevLett.104.030402} {\bibfield  {journal} {\bibinfo  {journal}
  {Phys. Rev. Lett.}\ }\textbf {\bibinfo {volume} {104}},\ \bibinfo {pages}
  {030402} (\bibinfo {year} {2010})}\BibitemShut {NoStop}%
\bibitem [{\citenamefont {Will}\ \emph {et~al.}(2016)\citenamefont {Will},
  \citenamefont {Park}, \citenamefont {Yan}, \citenamefont {Loh},\ and\
  \citenamefont {Zwierlein}}]{Will2016}%
  \BibitemOpen
  \bibfield  {author} {\bibinfo {author} {\bibfnamefont {S.~A.}\ \bibnamefont
  {Will}}, \bibinfo {author} {\bibfnamefont {J.~W.}\ \bibnamefont {Park}},
  \bibinfo {author} {\bibfnamefont {Z.~Z.}\ \bibnamefont {Yan}}, \bibinfo
  {author} {\bibfnamefont {H.}~\bibnamefont {Loh}}, \ and\ \bibinfo {author}
  {\bibfnamefont {M.~W.}\ \bibnamefont {Zwierlein}},\ }\href {\doibase
  10.1103/PhysRevLett.116.225306} {\bibfield  {journal} {\bibinfo  {journal}
  {Phys. Rev. Lett.}\ }\textbf {\bibinfo {volume} {116}},\ \bibinfo {pages}
  {225306} (\bibinfo {year} {2016})}\BibitemShut {NoStop}%
\bibitem [{\citenamefont {Guo}\ \emph {et~al.}(2018)\citenamefont {Guo},
  \citenamefont {Ye}, \citenamefont {He}, \citenamefont
  {Qu{\'{e}}m{\'{e}}ner},\ and\ \citenamefont {Wang}}]{Guo2018}%
  \BibitemOpen
  \bibfield  {author} {\bibinfo {author} {\bibfnamefont {M.}~\bibnamefont
  {Guo}}, \bibinfo {author} {\bibfnamefont {X.}~\bibnamefont {Ye}}, \bibinfo
  {author} {\bibfnamefont {J.}~\bibnamefont {He}}, \bibinfo {author}
  {\bibfnamefont {G.}~\bibnamefont {Qu{\'{e}}m{\'{e}}ner}}, \ and\ \bibinfo
  {author} {\bibfnamefont {D.}~\bibnamefont {Wang}},\ }\href {\doibase
  10.1103/PhysRevA.97.020501} {\bibfield  {journal} {\bibinfo  {journal} {Phys.
  Rev. A}\ }\textbf {\bibinfo {volume} {97}},\ \bibinfo {pages} {020501}
  (\bibinfo {year} {2018})}\BibitemShut {NoStop}%
\bibitem [{\citenamefont {Blackmore}\ \emph {et~al.}(2018)\citenamefont
  {Blackmore}, \citenamefont {Caldwell}, \citenamefont {Gregory}, \citenamefont
  {Bridge}, \citenamefont {Sawant}, \citenamefont {Aldegunde}, \citenamefont
  {Mur-Petit}, \citenamefont {Jaksch}, \citenamefont {Hutson}, \citenamefont
  {Sauer}, \citenamefont {Tarbutt},\ and\ \citenamefont
  {Cornish}}]{Blackmore2018}%
  \BibitemOpen
  \bibfield  {author} {\bibinfo {author} {\bibfnamefont {J.~A.}\ \bibnamefont
  {Blackmore}}, \bibinfo {author} {\bibfnamefont {L.}~\bibnamefont {Caldwell}},
  \bibinfo {author} {\bibfnamefont {P.~D.}\ \bibnamefont {Gregory}}, \bibinfo
  {author} {\bibfnamefont {E.~M.}\ \bibnamefont {Bridge}}, \bibinfo {author}
  {\bibfnamefont {R.}~\bibnamefont {Sawant}}, \bibinfo {author} {\bibfnamefont
  {J.}~\bibnamefont {Aldegunde}}, \bibinfo {author} {\bibfnamefont
  {J.}~\bibnamefont {Mur-Petit}}, \bibinfo {author} {\bibfnamefont
  {D.}~\bibnamefont {Jaksch}}, \bibinfo {author} {\bibfnamefont {J.~M.}\
  \bibnamefont {Hutson}}, \bibinfo {author} {\bibfnamefont {B.~E.}\
  \bibnamefont {Sauer}}, \bibinfo {author} {\bibfnamefont {M.~R.}\ \bibnamefont
  {Tarbutt}}, \ and\ \bibinfo {author} {\bibfnamefont {S.~L.}\ \bibnamefont
  {Cornish}},\ }\href {\doibase 10.1088/2058-9565/aaee35} {\bibfield  {journal}
  {\bibinfo  {journal} {Quantum Sci. Technol.}\ }\textbf {\bibinfo {volume}
  {4}},\ \bibinfo {pages} {014010} (\bibinfo {year} {2018})}\BibitemShut
  {NoStop}%
\bibitem [{\citenamefont {Ni}\ \emph {et~al.}(2010)\citenamefont {Ni},
  \citenamefont {Ospelkaus}, \citenamefont {Wang}, \citenamefont
  {Qu{\'e}m{\'e}ner}, \citenamefont {Neyenhuis}, \citenamefont {De~Miranda},
  \citenamefont {Bohn}, \citenamefont {Ye},\ and\ \citenamefont
  {Jin}}]{Ni2010}%
  \BibitemOpen
  \bibfield  {author} {\bibinfo {author} {\bibfnamefont {K.-K.}\ \bibnamefont
  {Ni}}, \bibinfo {author} {\bibfnamefont {S.}~\bibnamefont {Ospelkaus}},
  \bibinfo {author} {\bibfnamefont {D.}~\bibnamefont {Wang}}, \bibinfo {author}
  {\bibfnamefont {G.}~\bibnamefont {Qu{\'e}m{\'e}ner}}, \bibinfo {author}
  {\bibfnamefont {B.}~\bibnamefont {Neyenhuis}}, \bibinfo {author}
  {\bibfnamefont {M.}~\bibnamefont {De~Miranda}}, \bibinfo {author}
  {\bibfnamefont {J.}~\bibnamefont {Bohn}}, \bibinfo {author} {\bibfnamefont
  {J.}~\bibnamefont {Ye}}, \ and\ \bibinfo {author} {\bibfnamefont
  {D.}~\bibnamefont {Jin}},\ }\href {\doibase 10.1038/nature08953} {\bibfield
  {journal} {\bibinfo  {journal} {Nature}\ }\textbf {\bibinfo {volume} {464}},\
  \bibinfo {pages} {1324} (\bibinfo {year} {2010})}\BibitemShut {NoStop}%
\bibitem [{\citenamefont {Ye}\ \emph {et~al.}(2018)\citenamefont {Ye},
  \citenamefont {Guo}, \citenamefont {Gonz{\'{a}}lez-Mart{\'{i}}nez},
  \citenamefont {Qu{\'{e}}m{\'{e}}ner},\ and\ \citenamefont {Wang}}]{Ye2018}%
  \BibitemOpen
  \bibfield  {author} {\bibinfo {author} {\bibfnamefont {X.}~\bibnamefont
  {Ye}}, \bibinfo {author} {\bibfnamefont {M.}~\bibnamefont {Guo}}, \bibinfo
  {author} {\bibfnamefont {M.~L.}\ \bibnamefont
  {Gonz{\'{a}}lez-Mart{\'{i}}nez}}, \bibinfo {author} {\bibfnamefont
  {G.}~\bibnamefont {Qu{\'{e}}m{\'{e}}ner}}, \ and\ \bibinfo {author}
  {\bibfnamefont {D.}~\bibnamefont {Wang}},\ }\href {\doibase
  10.1126/sciadv.aaq0083} {\bibfield  {journal} {\bibinfo  {journal} {Sci.
  Adv.}\ }\textbf {\bibinfo {volume} {4}},\ \bibinfo {pages} {eaaq0083}
  (\bibinfo {year} {2018})}\BibitemShut {NoStop}%
\bibitem [{\citenamefont {Park}\ \emph {et~al.}(2017)\citenamefont {Park},
  \citenamefont {Yan}, \citenamefont {Loh}, \citenamefont {Will},\ and\
  \citenamefont {Zwierlein}}]{Park2017}%
  \BibitemOpen
  \bibfield  {author} {\bibinfo {author} {\bibfnamefont {J.~W.}\ \bibnamefont
  {Park}}, \bibinfo {author} {\bibfnamefont {Z.~Z.}\ \bibnamefont {Yan}},
  \bibinfo {author} {\bibfnamefont {H.}~\bibnamefont {Loh}}, \bibinfo {author}
  {\bibfnamefont {S.~A.}\ \bibnamefont {Will}}, \ and\ \bibinfo {author}
  {\bibfnamefont {M.~W.}\ \bibnamefont {Zwierlein}},\ }\href {\doibase
  10.1126/science.aal5066} {\bibfield  {journal} {\bibinfo  {journal}
  {Science}\ }\textbf {\bibinfo {volume} {357}},\ \bibinfo {pages} {372}
  (\bibinfo {year} {2017})}\BibitemShut {NoStop}%
\bibitem [{\citenamefont {DeMille}(2002)}]{DeMille2002}%
  \BibitemOpen
  \bibfield  {author} {\bibinfo {author} {\bibfnamefont {D.}~\bibnamefont
  {DeMille}},\ }\href {\doibase 10.1103/PhysRevLett.88.067901} {\bibfield
  {journal} {\bibinfo  {journal} {Phys. Rev. Lett.}\ }\textbf {\bibinfo
  {volume} {88}},\ \bibinfo {pages} {067901} (\bibinfo {year}
  {2002})}\BibitemShut {NoStop}%
\bibitem [{\citenamefont {Peter}\ \emph {et~al.}(2012)\citenamefont {Peter},
  \citenamefont {M\"uller}, \citenamefont {Wessel},\ and\ \citenamefont
  {B\"uchler}}]{Peter2012}%
  \BibitemOpen
  \bibfield  {author} {\bibinfo {author} {\bibfnamefont {D.}~\bibnamefont
  {Peter}}, \bibinfo {author} {\bibfnamefont {S.}~\bibnamefont {M\"uller}},
  \bibinfo {author} {\bibfnamefont {S.}~\bibnamefont {Wessel}}, \ and\ \bibinfo
  {author} {\bibfnamefont {H.~P.}\ \bibnamefont {B\"uchler}},\ }\href {\doibase
  10.1103/PhysRevLett.109.025303} {\bibfield  {journal} {\bibinfo  {journal}
  {Phys. Rev. Lett.}\ }\textbf {\bibinfo {volume} {109}},\ \bibinfo {pages}
  {025303} (\bibinfo {year} {2012})}\BibitemShut {NoStop}%
\bibitem [{\citenamefont {Yao}\ \emph {et~al.}(2013)\citenamefont {Yao},
  \citenamefont {Gorshkov}, \citenamefont {Laumann}, \citenamefont
  {L{\"{a}}uchli}, \citenamefont {Ye},\ and\ \citenamefont {Lukin}}]{Yao2013}%
  \BibitemOpen
  \bibfield  {author} {\bibinfo {author} {\bibfnamefont {N.~Y.}\ \bibnamefont
  {Yao}}, \bibinfo {author} {\bibfnamefont {A.~V.}\ \bibnamefont {Gorshkov}},
  \bibinfo {author} {\bibfnamefont {C.~R.}\ \bibnamefont {Laumann}}, \bibinfo
  {author} {\bibfnamefont {A.~M.}\ \bibnamefont {L{\"{a}}uchli}}, \bibinfo
  {author} {\bibfnamefont {J.}~\bibnamefont {Ye}}, \ and\ \bibinfo {author}
  {\bibfnamefont {M.~D.}\ \bibnamefont {Lukin}},\ }\href {\doibase
  10.1103/PhysRevLett.110.185302} {\bibfield  {journal} {\bibinfo  {journal}
  {Phys. Rev. Lett.}\ }\textbf {\bibinfo {volume} {110}},\ \bibinfo {pages}
  {185302} (\bibinfo {year} {2013})}\BibitemShut {NoStop}%
\bibitem [{\citenamefont {Kotochigova}\ and\ \citenamefont
  {DeMille}(2010)}]{Kotochigova2010}%
  \BibitemOpen
  \bibfield  {author} {\bibinfo {author} {\bibfnamefont {S.}~\bibnamefont
  {Kotochigova}}\ and\ \bibinfo {author} {\bibfnamefont {D.}~\bibnamefont
  {DeMille}},\ }\href {\doibase 10.1103/PhysRevA.82.063421} {\bibfield
  {journal} {\bibinfo  {journal} {Phys. Rev. A}\ }\textbf {\bibinfo {volume}
  {82}},\ \bibinfo {pages} {063421} (\bibinfo {year} {2010})}\BibitemShut
  {NoStop}%
\bibitem [{\citenamefont {Neyenhuis}\ \emph {et~al.}(2012)\citenamefont
  {Neyenhuis}, \citenamefont {Yan}, \citenamefont {Moses}, \citenamefont
  {Covey}, \citenamefont {Chotia}, \citenamefont {Petrov}, \citenamefont
  {Kotochigova}, \citenamefont {Ye},\ and\ \citenamefont
  {Jin}}]{Neyenhuis2012}%
  \BibitemOpen
  \bibfield  {author} {\bibinfo {author} {\bibfnamefont {B.}~\bibnamefont
  {Neyenhuis}}, \bibinfo {author} {\bibfnamefont {B.}~\bibnamefont {Yan}},
  \bibinfo {author} {\bibfnamefont {S.~A.}\ \bibnamefont {Moses}}, \bibinfo
  {author} {\bibfnamefont {J.~P.}\ \bibnamefont {Covey}}, \bibinfo {author}
  {\bibfnamefont {A.}~\bibnamefont {Chotia}}, \bibinfo {author} {\bibfnamefont
  {A.}~\bibnamefont {Petrov}}, \bibinfo {author} {\bibfnamefont
  {S.}~\bibnamefont {Kotochigova}}, \bibinfo {author} {\bibfnamefont
  {J.}~\bibnamefont {Ye}}, \ and\ \bibinfo {author} {\bibfnamefont {D.~S.}\
  \bibnamefont {Jin}},\ }\href {\doibase 10.1103/PhysRevLett.109.230403}
  {\bibfield  {journal} {\bibinfo  {journal} {Phys. Rev. Lett.}\ }\textbf
  {\bibinfo {volume} {109}},\ \bibinfo {pages} {230403} (\bibinfo {year}
  {2012})}\BibitemShut {NoStop}%
\bibitem [{\citenamefont {Deng}\ and\ \citenamefont {Yi}(2015)}]{Deng2015}%
  \BibitemOpen
  \bibfield  {author} {\bibinfo {author} {\bibfnamefont {Y.}~\bibnamefont
  {Deng}}\ and\ \bibinfo {author} {\bibfnamefont {S.}~\bibnamefont {Yi}},\
  }\href {\doibase 10.1103/PhysRevA.92.033624} {\bibfield  {journal} {\bibinfo
  {journal} {Phys. Rev. A}\ }\textbf {\bibinfo {volume} {92}},\ \bibinfo
  {pages} {033624} (\bibinfo {year} {2015})}\BibitemShut {NoStop}%
\bibitem [{\citenamefont {Li}\ \emph {et~al.}(2017)\citenamefont {Li},
  \citenamefont {Petrov}, \citenamefont {Makrides}, \citenamefont {Tiesinga},\
  and\ \citenamefont {Kotochigova}}]{Li2017}%
  \BibitemOpen
  \bibfield  {author} {\bibinfo {author} {\bibfnamefont {M.}~\bibnamefont
  {Li}}, \bibinfo {author} {\bibfnamefont {A.}~\bibnamefont {Petrov}}, \bibinfo
  {author} {\bibfnamefont {C.}~\bibnamefont {Makrides}}, \bibinfo {author}
  {\bibfnamefont {E.}~\bibnamefont {Tiesinga}}, \ and\ \bibinfo {author}
  {\bibfnamefont {S.}~\bibnamefont {Kotochigova}},\ }\href {\doibase
  10.1103/PhysRevA.95.063422} {\bibfield  {journal} {\bibinfo  {journal} {Phys.
  Rev. A}\ }\textbf {\bibinfo {volume} {95}},\ \bibinfo {pages} {063422}
  (\bibinfo {year} {2017})}\BibitemShut {NoStop}%
\bibitem [{\citenamefont {Hazzard}\ \emph {et~al.}(2014)\citenamefont
  {Hazzard}, \citenamefont {Gadway}, \citenamefont {Foss-Feig}, \citenamefont
  {Yan}, \citenamefont {Moses}, \citenamefont {Covey}, \citenamefont {Yao},
  \citenamefont {Lukin}, \citenamefont {Ye}, \citenamefont {Jin},\ and\
  \citenamefont {Rey}}]{Hazzard2014a}%
  \BibitemOpen
  \bibfield  {author} {\bibinfo {author} {\bibfnamefont {K.~R.~A.}\
  \bibnamefont {Hazzard}}, \bibinfo {author} {\bibfnamefont {B.}~\bibnamefont
  {Gadway}}, \bibinfo {author} {\bibfnamefont {M.}~\bibnamefont {Foss-Feig}},
  \bibinfo {author} {\bibfnamefont {B.}~\bibnamefont {Yan}}, \bibinfo {author}
  {\bibfnamefont {S.~A.}\ \bibnamefont {Moses}}, \bibinfo {author}
  {\bibfnamefont {J.~P.}\ \bibnamefont {Covey}}, \bibinfo {author}
  {\bibfnamefont {N.~Y.}\ \bibnamefont {Yao}}, \bibinfo {author} {\bibfnamefont
  {M.~D.}\ \bibnamefont {Lukin}}, \bibinfo {author} {\bibfnamefont
  {J.}~\bibnamefont {Ye}}, \bibinfo {author} {\bibfnamefont {D.~S.}\
  \bibnamefont {Jin}}, \ and\ \bibinfo {author} {\bibfnamefont {A.~M.}\
  \bibnamefont {Rey}},\ }\href {\doibase 10.1103/PhysRevLett.113.195302}
  {\bibfield  {journal} {\bibinfo  {journal} {Phys. Rev. Lett.}\ }\textbf
  {\bibinfo {volume} {113}},\ \bibinfo {pages} {195302} (\bibinfo {year}
  {2014})}\BibitemShut {NoStop}%
\bibitem [{\citenamefont {Gerdes}\ \emph {et~al.}(2011)\citenamefont {Gerdes},
  \citenamefont {Dulieu}, \citenamefont {Kn{\"{o}}ckel},\ and\ \citenamefont
  {Tiemann}}]{Gerdes2011}%
  \BibitemOpen
  \bibfield  {author} {\bibinfo {author} {\bibfnamefont {A.}~\bibnamefont
  {Gerdes}}, \bibinfo {author} {\bibfnamefont {O.}~\bibnamefont {Dulieu}},
  \bibinfo {author} {\bibfnamefont {H.}~\bibnamefont {Kn{\"{o}}ckel}}, \ and\
  \bibinfo {author} {\bibfnamefont {E.}~\bibnamefont {Tiemann}},\ }\href
  {\doibase 10.1140/epjd/e2011-20048-9} {\bibfield  {journal} {\bibinfo
  {journal} {Eur. Phys. J. D}\ }\textbf {\bibinfo {volume} {65}},\ \bibinfo
  {pages} {105} (\bibinfo {year} {2011})}\BibitemShut {NoStop}%
\bibitem [{\citenamefont {Steck}(2010)}]{Steck2003}%
  \BibitemOpen
  \bibfield  {author} {\bibinfo {author} {\bibfnamefont {D.~A.}\ \bibnamefont
  {Steck}},\ }\href {http://steck.us/alkalidata/sodiumnumbers.pdf} {\enquote
  {\bibinfo {title} {{Sodium D Line Data}},}\ } (\bibinfo {year}
  {2010})\BibitemShut {NoStop}%
\bibitem [{\citenamefont {Tiecke}(2011)}]{Tiecke2010}%
  \BibitemOpen
  \bibfield  {author} {\bibinfo {author} {\bibfnamefont {T.~G.}\ \bibnamefont
  {Tiecke}},\ }\href
  {{http://www.tobiastiecke.nl/archive/PotassiumProperties.pdf}} {\enquote
  {\bibinfo {title} {{Properties of Potassium}},}\ } (\bibinfo {year}
  {2011})\BibitemShut {NoStop}%
\bibitem [{Sup()}]{Supplement}%
  \BibitemOpen
  \href@noop {} {}\bibinfo {note} {See Supplemental Material for a discussion
  of lattice calibration, electric field setup, ac-Stark map, fitting bias of
  the contrast measurements, the MACE simulation and estimations of the dipolar
  interaction and the single particle dephasing.}\BibitemShut {Stop}%
\bibitem [{\citenamefont {Gregory}\ \emph {et~al.}(2017)\citenamefont
  {Gregory}, \citenamefont {Blackmore}, \citenamefont {Aldegunde},
  \citenamefont {Hutson},\ and\ \citenamefont {Cornish}}]{Gregory2017}%
  \BibitemOpen
  \bibfield  {author} {\bibinfo {author} {\bibfnamefont {P.~D.}\ \bibnamefont
  {Gregory}}, \bibinfo {author} {\bibfnamefont {J.~A.}\ \bibnamefont
  {Blackmore}}, \bibinfo {author} {\bibfnamefont {J.}~\bibnamefont
  {Aldegunde}}, \bibinfo {author} {\bibfnamefont {J.~M.}\ \bibnamefont
  {Hutson}}, \ and\ \bibinfo {author} {\bibfnamefont {S.~L.}\ \bibnamefont
  {Cornish}},\ }\href {\doibase 10.1103/PhysRevA.96.021402} {\bibfield
  {journal} {\bibinfo  {journal} {Phys. Rev. A}\ }\textbf {\bibinfo {volume}
  {96}},\ \bibinfo {pages} {021402} (\bibinfo {year} {2017})}\BibitemShut
  {NoStop}%
\bibitem [{\citenamefont {Ramsey}(1950)}]{Ramsey1950}%
  \BibitemOpen
  \bibfield  {author} {\bibinfo {author} {\bibfnamefont {N.~F.}\ \bibnamefont
  {Ramsey}},\ }\href {\doibase 10.1103/PhysRev.78.695} {\bibfield  {journal}
  {\bibinfo  {journal} {Phys. Rev.}\ }\textbf {\bibinfo {volume} {78}},\
  \bibinfo {pages} {695} (\bibinfo {year} {1950})}\BibitemShut {NoStop}%
\bibitem [{\citenamefont {Koller}\ \emph {et~al.}(2015)\citenamefont {Koller},
  \citenamefont {Mundinger}, \citenamefont {Wall},\ and\ \citenamefont
  {Rey}}]{Koller2015}%
  \BibitemOpen
  \bibfield  {author} {\bibinfo {author} {\bibfnamefont {A.~P.}\ \bibnamefont
  {Koller}}, \bibinfo {author} {\bibfnamefont {J.}~\bibnamefont {Mundinger}},
  \bibinfo {author} {\bibfnamefont {M.~L.}\ \bibnamefont {Wall}}, \ and\
  \bibinfo {author} {\bibfnamefont {A.~M.}\ \bibnamefont {Rey}},\ }\href
  {\doibase 10.1103/PhysRevA.92.033608} {\bibfield  {journal} {\bibinfo
  {journal} {Phys. Rev. A}\ }\textbf {\bibinfo {volume} {92}},\ \bibinfo
  {pages} {033608} (\bibinfo {year} {2015})}\BibitemShut {NoStop}%
\bibitem [{\citenamefont {Micheli}\ \emph {et~al.}(2007)\citenamefont
  {Micheli}, \citenamefont {Pupillo}, \citenamefont {B\"uchler},\ and\
  \citenamefont {Zoller}}]{Micheli2007}%
  \BibitemOpen
  \bibfield  {author} {\bibinfo {author} {\bibfnamefont {A.}~\bibnamefont
  {Micheli}}, \bibinfo {author} {\bibfnamefont {G.}~\bibnamefont {Pupillo}},
  \bibinfo {author} {\bibfnamefont {H.~P.}\ \bibnamefont {B\"uchler}}, \ and\
  \bibinfo {author} {\bibfnamefont {P.}~\bibnamefont {Zoller}},\ }\href
  {\doibase 10.1103/PhysRevA.76.043604} {\bibfield  {journal} {\bibinfo
  {journal} {Phys. Rev. A}\ }\textbf {\bibinfo {volume} {76}},\ \bibinfo
  {pages} {043604} (\bibinfo {year} {2007})}\BibitemShut {NoStop}%
\bibitem [{\citenamefont {Gorshkov}\ \emph {et~al.}(2011)\citenamefont
  {Gorshkov}, \citenamefont {Manmana}, \citenamefont {Chen}, \citenamefont
  {Demler}, \citenamefont {Lukin},\ and\ \citenamefont {Rey}}]{Gorshkov2011b}%
  \BibitemOpen
  \bibfield  {author} {\bibinfo {author} {\bibfnamefont {A.~V.}\ \bibnamefont
  {Gorshkov}}, \bibinfo {author} {\bibfnamefont {S.~R.}\ \bibnamefont
  {Manmana}}, \bibinfo {author} {\bibfnamefont {G.}~\bibnamefont {Chen}},
  \bibinfo {author} {\bibfnamefont {E.}~\bibnamefont {Demler}}, \bibinfo
  {author} {\bibfnamefont {M.~D.}\ \bibnamefont {Lukin}}, \ and\ \bibinfo
  {author} {\bibfnamefont {A.~M.}\ \bibnamefont {Rey}},\ }\href {\doibase
  10.1103/PhysRevA.84.033619} {\bibfield  {journal} {\bibinfo  {journal} {Phys.
  Rev. A}\ }\textbf {\bibinfo {volume} {84}},\ \bibinfo {pages} {033619}
  (\bibinfo {year} {2011})}\BibitemShut {NoStop}%
\bibitem [{\citenamefont {Kwasigroch}\ and\ \citenamefont
  {Cooper}(2014)}]{Kwasigroch2014}%
  \BibitemOpen
  \bibfield  {author} {\bibinfo {author} {\bibfnamefont {M.~P.}\ \bibnamefont
  {Kwasigroch}}\ and\ \bibinfo {author} {\bibfnamefont {N.~R.}\ \bibnamefont
  {Cooper}},\ }\href {\doibase 10.1103/PhysRevA.90.021605} {\bibfield
  {journal} {\bibinfo  {journal} {Phys. Rev. A}\ }\textbf {\bibinfo {volume}
  {90}},\ \bibinfo {pages} {021605} (\bibinfo {year} {2014})}\BibitemShut
  {NoStop}%
\bibitem [{\citenamefont {Friebel}\ \emph {et~al.}(1998)\citenamefont
  {Friebel}, \citenamefont {D'Andrea}, \citenamefont {Walz}, \citenamefont
  {Weitz},\ and\ \citenamefont {H{\"{a}}nsch}}]{Friebel1998}%
  \BibitemOpen
  \bibfield  {author} {\bibinfo {author} {\bibfnamefont {S.}~\bibnamefont
  {Friebel}}, \bibinfo {author} {\bibfnamefont {C.}~\bibnamefont {D'Andrea}},
  \bibinfo {author} {\bibfnamefont {J.}~\bibnamefont {Walz}}, \bibinfo {author}
  {\bibfnamefont {M.}~\bibnamefont {Weitz}}, \ and\ \bibinfo {author}
  {\bibfnamefont {T.~W.}\ \bibnamefont {H{\"{a}}nsch}},\ }\href {\doibase
  10.1103/PhysRevA.57.R20} {\bibfield  {journal} {\bibinfo  {journal} {Phys.
  Rev. A}\ }\textbf {\bibinfo {volume} {57}},\ \bibinfo {pages} {R20} (\bibinfo
  {year} {1998})}\BibitemShut {NoStop}%
\bibitem [{\citenamefont {Grimm}\ \emph {et~al.}(2000)\citenamefont {Grimm},
  \citenamefont {Weidem{\"{u}}ller},\ and\ \citenamefont
  {Ovchinnikov}}]{Grimm2000}%
  \BibitemOpen
  \bibfield  {author} {\bibinfo {author} {\bibfnamefont {R.}~\bibnamefont
  {Grimm}}, \bibinfo {author} {\bibfnamefont {M.}~\bibnamefont
  {Weidem{\"{u}}ller}}, \ and\ \bibinfo {author} {\bibfnamefont {Y.~B.}\
  \bibnamefont {Ovchinnikov}},\ }in\ \href {\doibase
  10.1016/S1049-250X(08)60186-X} {\emph {\bibinfo {booktitle} {Advances In
  Atomic, Molecular, and Optical Physics}}},\ Vol.~\bibinfo {volume} {42}\
  (\bibinfo  {publisher} {Academic Press},\ \bibinfo {year} {2000})\ pp.\
  \bibinfo {pages} {95--170}\BibitemShut {NoStop}%
\bibitem [{\citenamefont {H{\"{a}}ndel}(2010)}]{Handel2010}%
  \BibitemOpen
  \bibfield  {author} {\bibinfo {author} {\bibfnamefont {P.}~\bibnamefont
  {H{\"{a}}ndel}},\ }\href {\doibase 10.1016/j.measurement.2010.02.007}
  {\bibfield  {journal} {\bibinfo  {journal} {Measurement}\ }\textbf {\bibinfo
  {volume} {43}},\ \bibinfo {pages} {766} (\bibinfo {year} {2010})}\BibitemShut
  {NoStop}%
\bibitem [{\citenamefont {Shaw}(2015)}]{Shaw2015}%
  \BibitemOpen
  \bibfield  {author} {\bibinfo {author} {\bibfnamefont {J.}~\bibnamefont
  {Shaw}},\ }\href
  {https://scholar.colorado.edu/cgi/viewcontent.cgi?article=2136&context=honr_theses}
  {\enquote {\bibinfo {title} {{External Electric Fields: A New Tool for
  Controlling Ultracold Polar Molecules}},}\ } (\bibinfo {year}
  {2015})\BibitemShut {NoStop}%
\bibitem [{\citenamefont {Li}\ \emph {et~al.}(2013)\citenamefont {Li},
  \citenamefont {Kar},\ and\ \citenamefont {Jiang}}]{Na2013calculations}%
  \BibitemOpen
  \bibfield  {author} {\bibinfo {author} {\bibfnamefont {H.-W.}\ \bibnamefont
  {Li}}, \bibinfo {author} {\bibfnamefont {S.}~\bibnamefont {Kar}}, \ and\
  \bibinfo {author} {\bibfnamefont {P.}~\bibnamefont {Jiang}},\ }\href@noop {}
  {\bibfield  {journal} {\bibinfo  {journal} {Int. J. Quantum Chem.}\ }\textbf
  {\bibinfo {volume} {113}},\ \bibinfo {pages} {1493} (\bibinfo {year}
  {2013})}\BibitemShut {NoStop}%
\end{thebibliography}%

\newpage 
{\color{white}w}
\newpage 

\section*{Supplementary Information}
\subsection{Lattice calibration}
\begin{figure*}
\centering
\includegraphics{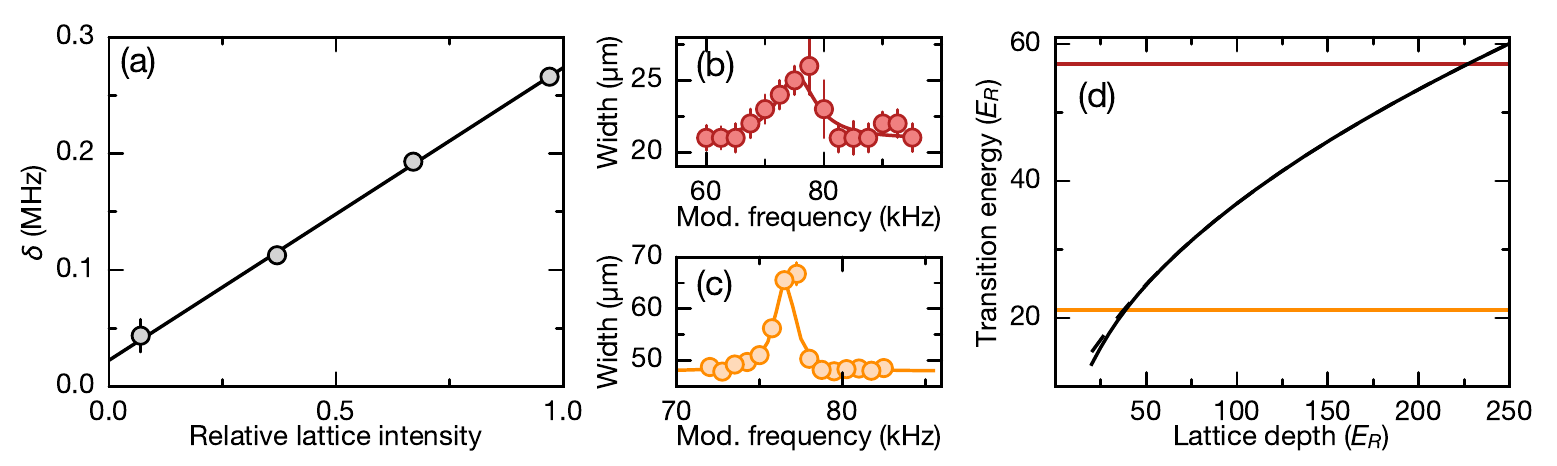}
\caption{\label{fig:intensity}
Lattice intensity calibration.
(a) Differential ac Stark shift between $\ket{\downarrow}$ and Feshbach molecular state  measured as STIRAP two-photon detuning $\delta$ for different lattice intensities.
Circles denote the center frequencies of Lorentzian fits to the spectra recorded at each intensity, error bars are derived from the covariance matrix of the fit.
The line is a linear fit to the center frequencies.
Parametric heating expansion measurement for \NaK molecules (b) and Na atoms (c).
Lines in (b) and (c) are Lorentzian fits to the data.
(d) Normalized transition frequency $f_{0\rightarrow2}$ as a function of the lattice depth when the quasi-momentum $q=0$ (black solid line) and $q=\hbar k$ (black dashed line), where $\hbar k$ is the recoil momentum of the lattice. 
The red (orange) line denotes the measured transition frequency of the molecules (Na atoms).
}
\end{figure*}

To determine the maximum intensity of the lattice $I_0$, we measure the differential ac Stark shift between the molecular $\ket{\downarrow}$ state and the Feshbach molecular state using two-photon STIRAP spectroscopy at varying lattice intensities $I$. 
The observed relationship between two-photon detuning $\delta$ and $I$ is linear, as can be seen in Fig. \ref{fig:intensity} a).
The slope of $251(4)$~kHz/$I_0$ amounts to the effective polarizability $\alpha_{\mathrm{eff}}$, the difference between polarizabilities of the initial Feshbach molecular state $\ket{FB}$ and the rovibronic ground state $\ket{\downarrow}$, which is given by 
\begin{equation}
\label{eq:intens}
\alpha_{\mathrm{eff}} = \alpha_{FB} - \alpha_{\ket{\downarrow}} = \alpha_{\mathrm{Na}} + \alpha_{\mathrm{K}} - \alpha_{\ket{\downarrow}}.
\end{equation}
The polarizability of Feshbach molecules $\alpha_{FB}$ is well approximated by the sum of the polarizabilities of the constituent sodium and potassium atoms, $\alpha_{\mathrm{Na}}$ and $\alpha_{\mathrm{K}}$. 
The atomic polarizabilities $\alpha_{\mathrm{Na}} = h\times9.0$~ Hz/(W/cm$^2$) \cite{Na2013calculations} and $\alpha_{\mathrm{K}}=h\times17.6$~Hz/(W/cm$^2$) \cite{Grimm2000} in a 1550~nm trap.  
$\alpha_{\ket{\downarrow}}$ is determined experimentally using parametric heating measurements \cite{Friebel1998, Neyenhuis2012, Gregory2017, Rvachov2017} of the lattice depth of molecules and sodium atoms, which are related by
\begin{equation}
\alpha_{\ket{\downarrow}}= \frac{V^{\mathrm{NaK}}_{\text{lat}}}{V^{\mathrm{Na}}_{\text{lat}}}\alpha_{\mathrm{Na}},
\end{equation}
where $V^{\mathrm{Na}}_{\text{lat}}$ ($V^{\mathrm{NaK}}_{\text{lat}}$) denotes the respective lattice depth of sodium atoms and molecules. 
In such a measurement the lattice intensity is modulated by $2.5\%$ for 8~ms.
Then the molecules (atoms) are released from the lattice and the cloud radius along $z$-direction is recorded, see Fig. \ref{fig:intensity} (b) and (c).  
Parametric heating occurs when the modulation frequency is equal to the transition frequency from the ground band to the second excited band, $f_{0\rightarrow2}$. 
We numerically solve the band structure of the optical lattice and obtain $f_{0\rightarrow2}$ as a function of the lattice depth, as shown in \ref{fig:intensity} d).
From our measurements we obtain $f_{0\rightarrow2}^{\mathrm{Na}}=76.5(2) $~kHz and $f_{0\rightarrow2}^{\mathrm{NaK}}= 75.2(8)$~kHz. 
The corresponding lattice depth is 38.6(2) $E_R^{\mathrm{Na}}$ for sodium atoms and 226(4) $E_R^{\mathrm{NaK}}$ for \NaK molecules respectively, where $E_R^{\mathrm{Na}}=h\times 3.612$~kHz and $E_R^{\mathrm{NaK}}=h\times 1.319$~kHz are the recoil energies of sodium atoms and \NaK molecules respectively.
We obtain $\alpha_{\ket{\downarrow}}=h\times19.3(4)$~Hz/(W/cm$^2$), which agrees well with the theoretical \textit{ab initio} value $\alpha^{\text{theory}}_{\ket{\downarrow}}=h\times20.4$~Hz/(W/cm$^2$).

With these polarizabilities, we can calculate $\alpha_{\mathrm{eff}}$ according to Eq. \ref{eq:intens} and determine the maximum lattice intensity to be $I_0=34(2)$~kW/cm$^2$.

\subsection{Electric field generation and calibration}
Direction, strength and gradient of the electric field are controlled with three groups of rod electrodes along the $x$, $y$ and $z$ axis. 
Each group consists of four parallel rods.

The in-vacuum main electrodes along the $x$-direction shown in Fig. \ref{fig:schematic}~a) generate a near homogeneous electric field, in this work pointing along the $y$-direction.  
The other two groups of auxiliary electrodes along $y$ and $z$ axis are mounted outside the glass cell, are set up in a quadrupole configuration.
They do not change the magnitude of the electric field at the position of the molecules to first order. 
We also add quadrupole voltages to the main electrodes and thus can compensate electric field gradients along all directions.
This is necessary due to inhomogeneities of the main electrodes and electric charge accumulation on the glass cell walls, which can only partially be removed with UV light in the beginning of each experimental cycle.

Each of the main electrodes is individually connected to a high precision voltage source with $\pm 400$~V with an rms noise of 0.55 mV from 10~\SI{}{\micro Hz} to 15 kHz \cite{Shaw2015}.
The voltage sources are controlled by high precision digital-to-analog converters (DACs).
The auxiliary electrodes are controlled by voltage sources with an rms noise of 1~mV. 
This allows us to apply stable electric fields up to $\pm 160$~V/cm to the molecules. 

\subsection{ac-Stark maps of the rotational transition}
We perform MW spectroscopy at two electric field strengths, see right panels in Fig. \ref{fig:acstark}. 
To ensure identical starting conditions for all data points, the preparation of molecules is always performed at the same lattice intensity, which is then ramped quickly to the respective spectroscopy intensity shortly before the MW sweep.
Afterwards the lattice is ramped back and the remaining $J=0$ molecules are detected as described in \cite{Seesselberg2018}.
Whenever $J=0$ molecules are lost, it is assumed that a transition to $J=1$ has occurred.

The theoretically expected frequency and strength for each MW transition are indicated by the lines in the left panels of Fig. \ref{fig:acstark}, where the strength of the transition is color coded. 
Only transitions with strengths larger than 0.5\% are displayed for clarity. 

In order to model the energies of all relevant rotational hyperfine levels in various external field set-ups, we follow the formalism of Ref.~\cite{Li2017} and the references therein. 
In the rovibronic ground state manifold the effective Hamiltonian includes interactions from rotation, hyperfine, Zeeman, ac and dc Stark effects. 
We evaluate the effective Hamiltonian in the zero-field rotational hyperfine basis with $J=0$ to $3$, which is then diagonalized to obtain the eigenenergies and eigenvectors at various external field settings. 
The same set of parameters to describe various interactions are used as in Ref.~\cite{Li2017} except dynamic polarizabilities at 1550~nm. 
We use the experimentally determined isotropic polarizability at 1550~nm, $\alpha_\mathrm{iso} = (2\alpha_{\perp}+\alpha_{\parallelsum})/3 = h\times19.3$ Hz/(W/cm$^2$), where $\alpha_{\perp}$ and $\alpha_{\parallelsum}$ are dynamic perpendicular and parallel radial electronic polarizabilities. 
The polarizability difference $\Delta\alpha_\mathrm{ele}= \alpha_{\parallelsum} - \alpha_\mathrm{\perp}$ is obtained by fitting the experimental transition frequencies $\nu$  at $E=144.3$~V/cm$^2$ and various laser intensities and polarization angles to $\Delta\alpha(\phi) = 2/15 \times (1-3\cos^2\phi)(\alpha_{\parallelsum} - \alpha_\mathrm{\perp})$.
The fitted values are $\alpha_{\perp}=h\times12$ Hz/(W/cm$^2$) and $\alpha_{\parallelsum}=h\times34$ Hz/(W/cm$^2$). 
Frank-Condon overlaps calculated from the eigenvectors are used as transition probabilities.

\subsection{Ramsey and spin-echo contrast bias estimation}
\begin{figure}
\centering
\includegraphics{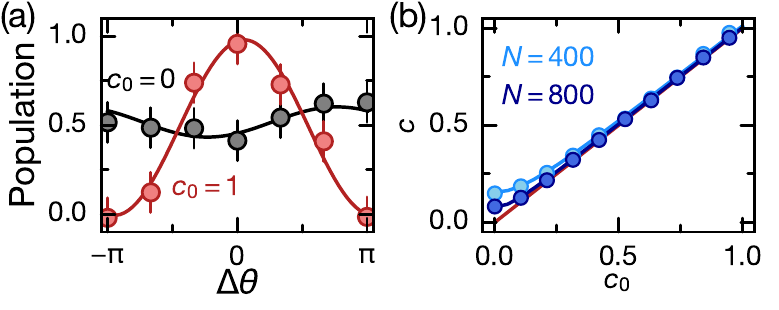}
\caption{(a) Simulated interference fringes in presence of molecule number noise when $N_{\mathrm{tot}}=400$. 
Symbols are the simulated population as a function of $\Delta\theta$, solid lines are the fitted interference fringes. 
The errorbar is the error of the mean of the molecule number. 
(b) Measured contrast $c$ as function of the real contrast $c_0$ when $N_{\mathrm{tot}}=$400 (light blue) or 800 (dark blue). 
Solid lines in the same color are the corresponding contrasts given by Eq.~\eqref{eqs2}. 
The red line gives the measured contrast without noise. 
In all simulations $\Delta N=98$, and $M_{s}=35$ which is similar as in experiments.}
\label{ContrastBias}
\end{figure}

\begin{figure}
\centering
\includegraphics[width=0.5\textwidth]{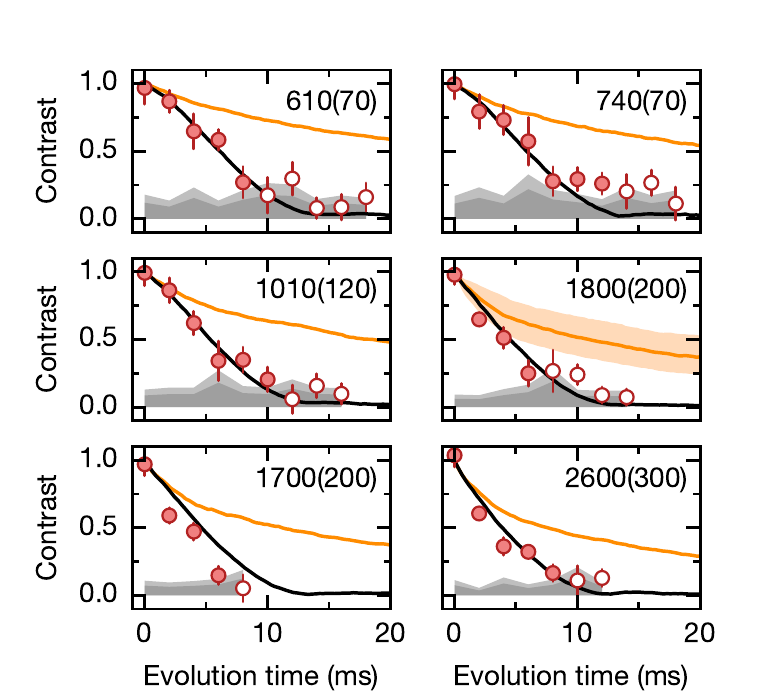}
\caption{\label{FigS1}
Comparison of measurement (circles) and simulation (lines) of the Ramsey experiments. 
Numbers in the right corner denote the initial molecule number(error).
Symbols are measured contrast values as function of evolution time. 
Black (orange) solid lines are MACE simulations with (without) external field gradient $h\times1.3(1)$~Hz/\SI{}{\micro m}. 
The orange shaded region marks the effect of a factor of two change in density in either direction on the simulation. 
The dark grey shaded region indicates the bias of the data, the light grey region the cutoff for the data (empty circles).}
\end{figure}

To avoid ambiguities in the Ramsey fringe fitting due to an unknown and potentially slowly drifting phase of the fringe, we restrict the contrast of the fringe $c$ to positive values only.
However, due to molecule number fluctuations, the fit contrast $c$ of an experimental Ramsey fringe is never 0.
This biases $c$, especially when the molecule number is small~\cite{Handel2010}. 
To quantitatively understand this, we simulate Ramsey interference fringes in presence of molecule number fluctuations, see Fig. \ref{ContrastBias}. 
We add Gaussian noise with a standard deviation $\Delta N$ to the molecule number of an ideal sinusoidal fringe with contrast $c_0$ and fit the resulting contrast. 
We repeat this simulation 300 times to obtain an average measured contrast $c$ and find that the contrast bias $\Delta c$ adds to the real contrast quadratically as
\begin{align}
c =\sqrt{c^2_0+\Delta c^2} \label{eqs2},
\end{align}
where
\begin{equation}
\Delta c =\sqrt{\frac{a}{M_{s}}}\frac{2\Delta N}{N_{\mathrm{tot}}}. 
\end{equation}
$M_{s}$ is the sampling size and $a=3.5$ is an empirical parameter obtained from our simulations. 
In our experiments, $\Delta c \approx 10\%$ depending on the molecule number, see Fig. \ref{FigS1} and Fig. \ref{FigS2} (dark grey shaded areas). 
To avoid overestimation of the coherence time, we exclude data (empty circles) with a contrast smaller than $1.5\Delta c$ (light grey shaded areas) for fitting. 
The remaining data is fit with a Gaussian function of the form
\begin{equation}
c(t) = c_i \exp\left(-\left(\frac{t}{\tau}\right)^2\right),
\label{eq:gauss}
\end{equation}
where $c_i$ is the initial contrast at the shortest evolution time and $\tau$ denotes the coherence time.

Another approach would be to subtract the contrast bias using Eq.~\eqref{eqs2} before the fitting.
We found that in this case the overestimation is below $10\%$, even when the low contrast data is included. 
This is because the the tails of the Gaussian decay curve in Eq. \eqref{eq:gauss} contribute less to the fitting than the high contrast data. 

The detection offset of the molecule number is less than 20 molecules and thus negligible.

\subsection{Estimation of dipolar interaction}
The mean vaule of the angular-independent part of $U_{ij}$ sets an energy scale for the dipolar interaction, which is given as
\begin{equation}
\label{eq:avint}
\langle U_d \rangle \approx \frac{2|d^2_ {\uparrow\downarrow}|}{4\pi\varepsilon_0 l^3},
\end{equation}
where $l=n_0^{-1/3}$ is the average distance between molecules  and $n_0$ is the peak molecular density.

As the lattice spacing 0.775 \SI{}{\micro m} is much less than the average distance between molecules, the density distribution of the molecular gas is approximated by 
\begin{equation}
n(x,y,z)=n_0\exp\left(-\frac{x^2}{\sigma_x^2}-\frac{y^2}{\sigma_y^2}-\frac{z^2}{\sigma_z^2}\right), 
\end{equation}
where the peak molecular density is
\begin{equation}
n_0=\frac{N_{\mathrm{tot}}}{\pi^{3/2}\sigma_x\sigma_y\sigma_z}, 
\end{equation}
where $\sigma_x=\sigma_y=$ 27(4) \SI{}{\micro m} and $\sigma_z=$11.5(6) \SI{}{\micro m} are the $1/e$ radius of the molecular cloud in $x$-, $y$- or $z$-direction as determined by \textit{in situ} imaging. 
In the experiment we change the molecule number by varying the hold time between Feshbach molecule production and further experiments, which leads to loss according to inelastic collisions.
This allows us to keep the cloud radius almost independent of the molecule number. 
The peak density for the highest molecule number 3200(300) is $7(3)\times 10^{10}$~/cm$^3$ and the corresponding average distance is 2.4(3) \SI{}{\micro m}.
This results in a peak dipolar interaction of $\langle U_d \rangle=h\times 50(20)$~Hz, similar to the decoherence rate we observe at the highest molecule number.

\subsection{Estimation of single-particle dephasing rates}
The transition frequency between $\ket{\downarrow}$ and $\ket{\uparrow}$  depends on temporal fluctuations and spatial variations in the external potential across the molecule cloud. 
These changes can be described by the MW detuning term $\Delta(\mathbf{r},t)$ of Eq. \ref{eqs1}, which can be written as 
\begin{equation}
\Delta(\mathbfit{r}, t)= \Delta\alpha I(\mathbfit{r})+\beta I^2(\mathbfit{r})+\xi E(\mathbfit{r},t)^2.
\label{eq:effmwd}
\end{equation}
The first two terms are first-order and second-order differential light shifts from Eq. \ref{eq:beta}.  
We assume that the lattice beam has a Gaussian intensity profile and ignore the intensity variation along z-direction
\begin{equation}
I(\mathbfit{r})=I_{\text{peak}}\exp \left(-2\frac{x^2+y^2}{\omega^2_0} \right)
\end{equation}
with beam waist $\omega_0=100$~\SI{}{\micro m}. 
The third term of Eq. \ref{eq:effmwd} is the differential dc Stark shift, where according to \cite{Micheli2007}
\begin{equation}
\xi=\frac{4}{15}\frac{d^2_0}{B_{\text{rot}}}=h\times177~\frac{\mathrm{Hz}}{\mathrm{(V/cm)}^2}.
\end{equation}
The electric field $E$ can be written as
\begin{equation}
E(\mathbfit{r},t)\approx E_0(t)+\nabla E\cdot \mathbfit{r}+\mathcal{O}(r^2), 
\end{equation}
where $E_0(t)$ describes the temporal fluctuation and the second term is the first-order inhomogeneity of $E$. 

The inhomogeneity of $\Delta(\mathbf{r})$ leads to dephasing of rotational excitations and can be calculated from Eq. \ref{eq:effmwd} by numerical integration.
The experimentally observed dephasing $\gamma$ thus has local and temporal contributions:
Due to residual differential light shifts, $\gamma_{\text{L}}$, gradient electric fields, $\gamma_{\text{EG}}$, and temporal fluctuations of $E$, $\gamma_{\text{EN}}$.
\begin{equation}
\gamma_{\text{L}}=\frac{\int{n(\mathbfit{r})\left|\delta\alpha \Delta I(\mathbfit{r})+\beta \Delta I^2(\mathbfit{r})\right|}\mathbf{d}\mathbfit{r}}{\int{n(\mathbfit{r})\mathbf{d}\mathbfit{r}}}
\end{equation}
is the dephasing due to residual differential light shifts, where $\delta\alpha=h (\partial\nu/\partial I)|_{I=I_{\text{peak}}}=\Delta\alpha +2 \beta I_{\text{peak}}$ is the local differential polarizability, 
$\Delta I(\mathbfit{r})=I(\mathbfit{r})-I_{\text{peak}}$, and $n(\mathbf{r})$ is the density described in the last section.
The dephasing rate due to gradient electric fields $|\nabla E|$ is given as $\gamma_{\text{EG}}=2\xi \overline{E_0}|\nabla E|\sigma_x$ assuming that the gradient is along x-direction for simplicity, where $\overline{E_0}$ is the time averaged electric field in the center of the molecular cloud.
The dephasing due to temporal electric field noise $\delta E_0$ is $\gamma_{\text{EN}}=2\xi\overline{E_0}\delta E_0$.
\begin{table}
\begin{center}
\begin{tabular}{ lll } 
   source & value & dephasing rate \\ 
   \hline
	$\delta\alpha$ & $ h\times 0.05~\text{Hz/(W/cm}^2)$ & \multirow{2}{*}{$\gamma_\text{L}=h\times$32~Hz} \\
   $\beta$ & $h\times 30~\text{Hz/(kW/cm}^2)^2$ & {} \\     
   $\delta E$ & $$0.5~mV/cm & $\gamma_\text{EN}=h\times$12~Hz\\ 
   $ \nabla  E$ & $$0.5~$\text{V/cm}^2$ & $\gamma_\text{EG}=h\times$38~Hz \\ 
\end{tabular}
\caption{Maximal dephasing rates at 68.3 V/cm due to imperfect cancellation of local differential polarizability $\delta\alpha$ and residual hyperpolarizability $\beta$, and due to electric field noise and gradients, $\delta E$ and $\nabla E$.}
\label{Tab1}
\end{center}
\end{table}
The estimated maximal dephasing rates are summarized in Tab. \ref{Tab1} together with their experimental origin. 

The effective dephasing rate can then be calculated as $\gamma=\sqrt{\gamma^2_{\text{L}}+\gamma^2_{\text{EG}}+2\gamma^2_{\text{EN}}}$ and is related to the coherence time by $\tau\approx 2\hbar/\gamma$ which is verified by numerical simulation.

The dephasing rate is on the order of a few ten Hz in the current setup. 
In the future, the dephasing rate could be reduced to a few Hz by implementing less noisy, more homogeneous dc electric field as well as more precise laser polarization control.

\subsection{MACE simulation of spin dynamics}
To understand the decoherence induced by the dipolar interaction, we implement the MACE simulation \cite{Hazzard2014a} in which molecules are spatially frozen and randomly distributed with a Gaussian probability distribution in tens of layers with a spacing of 0.775 \SI{}{\micro m}. 
The cloud radius along x-, y-, and z-direction in the simulation is the same as in the experiments.
In the simulation, we divide the molecules into hundreds of clusters of four molecules with the strongest dipolar interactions. 
Then we exactly solve the time evolution for each cluster and sum up the expectation values of all spins to obtain the Ramsey signal. 

For a homogeneous external field, the decoherence is due to the random spread of the dipolar interaction of molecules. 
We expect the coherence time $\tau_c$ to be inversely proportional to the molecule number $N_{\mathrm{mol}}$ because the dipolar interaction is proportional to the molecular density. 
In this homogeneous MACE simulation the 1/$e$ coherence time is about 12~ms with 3000 and 70~ms with 500 molecules, see orange lines in Fig. \ref{FigS1}. 
In the Ramsey experiments (circles) however, the coherence time is limited to about 8~ms even for 600 molecules when the dipolar interaction is negligible. 
This is due to the residual single particle dephasing discussed in the previous section. 
In order to qualitatively introduce this dephasing into the model, we implement a simple \textit{effective} external field gradient along x-direction $\Delta=\Delta' x$. 
By fitting the experimental data with lowest molecule number we obtain $\Delta'= h\times1.3(1)$~Hz/\SI{}{\micro m} which corresponds to a dephasing rate of $h\times35(2)$~Hz. 
Taking the effective dephasing into account, we can reproduce the experiments with various molecule numbers in the simulation (black lines).
\begin{figure}
\centering
\includegraphics[width=0.5\textwidth]{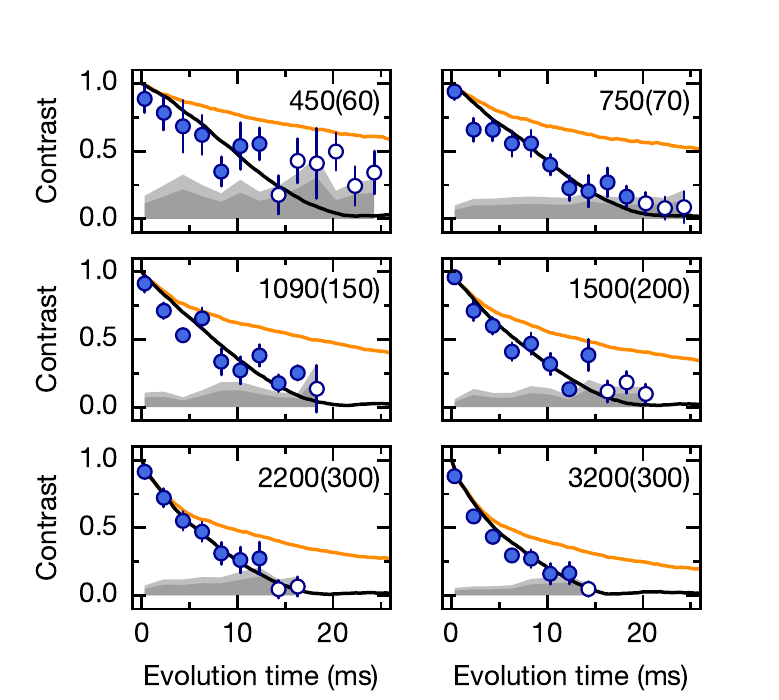}
\caption{\label{FigS2}
Comparison of measurement and simulation of the spin-echo experiments. 
As Fig. \ref{FigS1}, but with spin-echo pulse sequence and an external field gradient of $h\times0.8(1)$~Hz/\SI{}{\micro m} in the MACE simulation.}
\end{figure}
For the spin-echo experiments, see Fig. \ref{FigS2}, we obtain an effective external field gradient of $\Delta'=h\times 0.8(1)$~Hz/\SI{}{\micro m} in a similar manner to account for the uncancelled single particle dephasing of $h \times 21(2)$~Hz.
Implementing a parabolic external field in the simulation also produced similar decoherence behavior.

\end{document}